\documentclass[11pt,twocolumn]{article}

\usepackage{epsf}
\usepackage{cite} 

\def\s{\sigma}
\def\sgn{\mathop{\rm sign}\nolimits}
\def\half{{\textstyle{{1}\over{2}}}}
\def\df{\partial}
\def\nn{\nonumber}

\def\Z{{\bf Z}}

\def\mod{\mathop{\mbox{mod}}\nolimits}

\def\ket#1{\bigl|#1\bigr>}
\def\bra#1{\bigl<#1\bigr|}
\def\vsp{\vspace{-1mm}}
\def\ll{\mathop{<\!<}}

\def\quart{{\textstyle{{1}\over{4}}}}
\def\intt{\int_{0}^{2\pi}d\s}
\def\sqp{\sqrt{P^{2}}}
\def\astar{a^{*}}
\def\nm{n_{\mu}^{-}}
\def\q{Q_{\mu}(\sigma)}
\def\kr{\Gamma _{\mu}^{ij}=N_{\nu}^{i}\partial N_{\nu}^{j}/
   \partial \p}  
\def\p{P_{\mu}}
\def\sbo{{\vec S}}

\newlength{\mywidth}\mywidth=2.3truein 
\epsfysize=\mywidth
\epsfxsize=\mywidth
\def\fref#1{fig.\ref{#1}}

\renewenvironment{figure}{\refstepcounter{figure}
\baselineskip=0.4\normalbaselineskip\footnotesize}
{\baselineskip=\normalbaselineskip}
\def\fignum{{\bf Fig.\arabic{figure}.\quad}}
\renewenvironment{table}{\refstepcounter{table}}{}
\def\tabnum{Table \arabic{table}}

\begin{document}

\title{{\bf\large
Quantization of Non-Critical Bosonic Open String Theory
}}
\author{\large 
Igor Nikitin\\
{\it\normalsize 
Fraunhofer Society, IMK.VE, 53754, St.Augustin, Germany
}
}
\date{}
\maketitle

~

\vspace{-5mm}

\baselineskip=0.4\normalbaselineskip\footnotesize

The theory of relativistic strings is considered in frames 
of Hamiltonian formalism and Dirac's quantization procedure.
A special gauge fixing condition is formulated, 
related with the world sheet of the string in Lorentz-invariant way. 
As a result, a new set of Lorentz-invariant canonical variables 
is constructed, in which a consistent quantization
of bosonic open string theory could be done in Minkowski 
space-time of dimension $d=4$. The obtained quantum theory 
possesses spin-mass spectrum with Regge-like behavior.

\vspace{2mm}

\baselineskip=\normalbaselineskip\normalsize

\vsp\section{Introduction}

The model of relativistic string has been proposed 
by Nambu, Hara and Goto \cite{disref9a,disref9b,disref9c} 
in the early 1970s for the description of the internal structure
of strongly interacting particles, the hadrons.
The initial purpose was the explanation of spin-mass
spectra of the hadrons, as well as certain experimentally 
established properties of their interaction amplitudes 
\cite{disref10a,disref10b,disref10c}.
Nowadays the construction of the string-inspired models of the hadrons 
is continued \cite{NambuHiggs,chrom1,chrom2,Artru,thickness,
straight-em1,3str,3strSharov}. 
The idealized theoretical model was equipped by necessary 
physical details: quarks at the ends of the string, 
carrying masses, electric charges, spin and other
quantum numbers; different mechanisms of the string breaking, 
responsible for the decays of the hadrons. The progress of this construction
has been reported in the book \cite{BarbNest}. Meanwhile
the development of string theory selected a different way.
It was early recognized \cite{Hennaux-ref17,Hennaux-ref18,
Hennaux-ref18-2,Hennaux-ref18a}
that quantization of Nambu-Goto model possesses anomalies
at the physical dimension of the space-time $d=4$,
and is free of anomalies at $d=26$.
It was also noticed that inclusion of additional 
fermionic degrees of freedom to the theory \cite{disref26a,disref26b}
cancels the anomaly at less value of dimension $d=10$.
Later this approach was combined with the other idea:
some of dimensions were considered as coordinates on a compact
manifold of physically small size. 
Further developments, excellently reviewed in 
\cite{GrinSchwarzWitten,BrinkHennaux}, have included more complex 
mathematical structures to the theory, considered multi-dimensional 
extensions of the strings \hbox{(p-branes)}, as well as their supersymmetric 
analogs, and formed a powerful direction of modern theoretical physics, 
providing a basis for the construction of {\it Grand Unification Theory}. 
However, this approach cannot be
immediately applied in the models of hadrons, where the subject
of consideration is bosonic 4-dimensional Nambu-Goto string, 
while the introduction of extra dimensions and additional degrees 
of freedom changes this system essentially. 

Nowadays a lot of work has been done in the field
of {\it non-critical string theory}, studying the possibilities
for string quantization beyond the critical dimension $d=26$. 
Polyakov \cite{Polyakov} used the path integration technique
to construct the quantum string theory at arbitrary $d$.
It has been shown in this paper that at $d=26$ the theory 
admits oscillator representation, while at $d\neq 26$ the theory 
becomes equivalent to a non-linear field theory.
In operator formalism the problem was considered by 
Rohrlich \cite{Rohrlich}. In this paper a consistent quantum theory 
was constructed for Nambu-Goto string at arbitrary $d$.
For this purpose a special Lorentz-invariant parametrization 
of the world sheet (time-like gauge) was used, with the gauge axis 
directed along total momentum $P_{\mu}$. Lorentz-invariant
parametrizations were also used in papers
\cite{zone,slstring,straight-em1,ax,2par}
presenting other approaches to quantization of 
non-critical string theory.

This paper extends the approaches of 
\cite{Rohrlich,zone,slstring,straight-em1,2par,ax}
to the gauge of a special form, which can be considered
as Lorentz-invariant modification of light cone one.
The paper is organized as follows.
Section~2 introduces general concepts and a formalism,
used for the elimination of anomalies from the Lorentz group 
in string theory. This formalism is a combination of general methods, 
comprehensively described in \cite{zone,Dirac,Arnold-mat-phys}, 
which in fact can be applied to any theory. Wherever possible,
we provide the mathematical details of the methods and
the description of their physical meaning. In Sections~3 and 4 
we investigate the algebraic and geometric properties of 
the constructed mechanics. Section~5 is devoted to its quantization.
The obtained results are discussed in Section~6.

\vspace{-1mm}
\section{General concepts}\label{gen-concepts}

{\it Quantization} \cite{Dirac,geom-quant} is a linear mapping 
$f\to \hat f$ of a pre-defined set of classical variables 
to a set of operators, satisfying the following 
correspondence principle: $i\hbar\{f,g\}\to[\hat f,\hat g]$.
Here $\{,\}$ is Poisson bracket, $[,]$ is commutator,
$\hbar$ is Planck's constant. Additionally it is required 
that real-valued dynamical variables should be represented 
by Hermitian operators, the unity should be preserved: $1\to\hat1$,
the space of states should be positively defined and
irreducible under the action of the constructed operators 
(does not have a smaller subspace invariant under their action).

It is shown in \cite{geom-quant} that the correspondence principle 
cannot be satisfied for all dynamical variables in the theory. 
This property is related with ordering ambiguities 
for non-commuting operators. Indeed, let's consider two sets
of classical variables, related by some non-linear transformation:
$a_{i}\leftrightarrow b_{i}$. If the correspondence principle 
is satisfied for the set $a_{i}$: $\{a_{i},a_{j}\}\to[\hat a_{i},\hat a_{j}]$,
then the commutator of operators $[\hat b_{i}(\hat a),\hat b_{j}(\hat a)]$,
represented as non-linear functions of $\hat a$, can contain 
ordering ambiguities, which create anomalous terms
absent in Poisson bracket $\{b_{i},b_{j}\}$. In spite of
the fact that a contribution of such terms is suppressed
by additional factor $\hbar$, their occurrence can lead to
serious problems in quantum theory, especially if $b_{i}$ 
represent the generators of symmetries of the system.
On the other hand, if the correspondence principle 
is satisfied for the set $b_{i}$, the commutator 
$[\hat b_{i},\hat b_{j}]$ is postulated directly from 
Poisson bracket $\{b_{i},b_{j}\}$ and has no anomalies. 
The anomalies can appear
in the commutator $[\hat a_{i}(\hat b),\hat a_{j}(\hat b)]$.
The common practice is to select a convenient basis of independent
variables, where the quantization is performed, trying to keep all 
generators of symmetries simple functions of independent variables
and thus to avoid the occurrence of anomalies in their commutators.

To make this general consideration more concrete,
let $a_{i}$ represent the set of oscillator variables
of string theory in the light cone gauge, and
$b_{i}=(M_{\mu\nu},\xi_{i})$, where $M_{\mu\nu}$
are the generators of Lorentz group and $\xi_{i}$
is a set of variables complementing $M_{\mu\nu}$
to the full phase space. In selection of such a set
one needs to take care that $\xi_{i}$ will have
simple Poisson brackets with $M_{\mu\nu}$ and
among themselves. Local existence of such variables
is provided by Darboux theorem \cite{Arnold-mat-phys}, while
the determination of the global structure for their region 
of variation is generally a complex task \cite{zone}.
Further, performing the quantization of string theory
in variables $a_{i}$, we obtain well known anomaly
in the commutator $[M_{\mu\nu}(a),M_{\rho\s}(a)]$.
On the other hand, performing the quantization of string theory
in variables $b_{i}$, we do not have the anomaly in $[M_{\mu\nu},M_{\rho\s}]$.
Anomalies can appear in the commutator $[a_{i}(b),a_{j}(b)]$. 
However, the point is that 
there is no more necessity to use the variables $a_{i}$
in the theory, because the role of internal variables
is now played by $\xi_{i}$, whose commutator does not
have the anomalies. Also, if such necessity 
would appear, the anomaly in $[a_{i}(b),a_{j}(b)]$ 
does not lead to any further problems in the theory. 
Even if the anomalies would be excluded both from 
$[a_{i},a_{j}]$ and $[b_{i},b_{j}]$, one can always
find the third set of variables $c_{i}$, highly non-linear
in terms of independent ones that their commutator 
$[c_{i},c_{j}]$ will have the anomalies.
As we already mentioned,
the correspondence principle cannot be satisfied for all 
variables in the theory, according to the theorem \cite{geom-quant}.

It becomes clear that the anomalies essentially depend 
on the choice of canonical variables in which the quantization 
is performed. This property was previously discussed in \cite{zone}. 
Actually, this fact is well known 
and even presented in the textbooks on string theory, 
e.g. in \cite{BrinkHennaux} on p.157: 
``Should one not try to use a different representation
of the string operators so as to avoid the central charge?
Again, it might very well be possible to construct such 
a representation and, if so, it is very likely that the
resulting quantum theory would be very different from the one
explained here. It could be that this yet-to-be-constructed
theory would possess an intrinsic interest of its own
(e.g., through the occurrence of infinite-dimensional
representations of the Lorentz algebra). Moreover,
because this theory would not be based on the use of
oscillator variables, it might be more easily extendable
to higher-dimensional objects, such as the membrane.''
However, an opinion can be often 
encountered that anomalies in string theory are stable, 
i.e. can be transferred from one sector to the other one, 
but cannot be completely removed. Really this \hbox{no-go} theorem 
has never been proven. It is only proven that string theory in the
oscillator representation, formed by Fourier coefficients
of the world sheet expansion, has similar anomalies
both in covariant and non-covariant approach. It is not proven
that anomaly is present in every possible representation.
The paper \cite{Bowick} can be considered as an attempt
to build such a proof using geometrical quantization technique.
The geometrical quantization \cite{geom-quant} is an
implementation of canonical procedure by means of 
differential geometrical constructions available in 
the classical theory. Most of the elements of this technique
are coordinate independent. However, it was pointed out in
\cite{geom-quant} that as a whole it is not 
coordinate independent due to the choice of irreducible
component in the space of states, known also as polarization.
Holomorphic polarization chosen in \cite{Bowick}
makes the theory equivalent to the oscillator one,
reproducing the anomaly inherent to this representation.
Again, it has not been proven that the anomaly is present
in every polarization. It gradually becomes clear that
the above mentioned no-go theorem is not satisfied,
because the canonical quantization in the variables 
$(M_{\mu\nu},\xi_{i})$ immediately excludes the anomaly 
from the Lorentz group. Such a set was explicitly constructed 
in the paper \cite{zone}. An alternative set will be constructed 
in the given paper using a different representation.

Quantization of theories with gauge symmetries
has own peculiarities. On the classical level such theories
have Dirac's constraints $L_{n}$, which generate gauge
symmetries via Poisson brackets \cite{Dirac}. 
There are two main approaches for quantization of such theories.
In the first one, called covariant quantization, 
the constraints are imposed on the state vectors
$L_{n}\Psi=0$, thus selecting gauge invariant wave functions
in the space of states. If the algebra $[L_{n},L_{m}]$
has the anomaly, it cannot be implemented by means of
canonical formalism, but its certain extensions can be used,
such as implementation only a half of constraints 
in the theory \cite{BrinkHennaux}. In the second approach 
the gauge fixing conditions are imposed in addition 
to the primary constraints, which select one representative
from each gauge orbit. In this case the whole set of constraints 
belongs to the 2nd class in Dirac's terminology \cite{Dirac}, 
the mechanics should be reduced to their surface 
by corresponding redefinition of Poisson brackets,
and after that can be quantized. An intermediate position
is taken by the ``reduced phase space'' formalism \cite{nogauge},
where the role of the phase space is played by the factor-space
with respect to the group of gauge symmetry. 
In frames of this formalism the gauge fixation corresponds
to a choice of basis in the space of gauge invariants.
Indeed, in string theory the world sheet can be parameterized
by a temporal component of coordinate along some axis $n_{\mu}$.
Light-like vector $n_{\mu}$ corresponds to the light cone gauge,
while $n_{\mu}=P_{\mu}/\sqrt{P^{2}}$, where $P_{\mu}$
is total momentum, is Rohrlich's gauge \cite{Rohrlich}.
The oscillator variables are Fourier coefficients
$$a_{k}=\int d\s\; Q'(\s)e^{ik\s},$$ where $Q'(\s)=x'(\s)+p(\s)$
and $\s$ is the light cone or Rohrlich's gauge parameter:
$\s=\pi(nQ)/(nP)$. Obviously, the same variables can be written
in the form of parametric invariants $$a_{k}=\int dQ\;e^{ik\pi(nQ)/(nP)}.$$
For the light cone gauge this formula 
gives well known DDF variables \cite{DDF}. 
In terms of these variables the world sheet
can be reconstructed (up to translations). 
Thus, the gauge fixation does not make the theory 
restricted or weak in some other sense, but provides the particular
choice of basis in the space of gauge invariants. When $n_{\mu}$ 
is changed, the basis $a_{k}(n_{\mu})$ is changed accordingly. 
As we already know, such change influences the anomalies, 
which therefore depend on the selected gauge. 

This effect can be also understood using the following
geometrical consideration. In the standard light cone gauge 
the vector $n_{\mu}$ is non-dynamical, e.g.
$n_{\mu}=(1,1,0,0)$. Because the Lorentz transformations 
change the position of the world sheet with respect to this axis, 
they are followed by reparametrizations of the world sheet.
On quantum level the reparametrization group has anomaly,
which appears also in Lorentz group and violates
Lorentz covariance of the theory. This is a main problem
of string theory in standard light cone gauge.
On the other hand, the Rohrlich's gauge relates $n_{\mu}$
with the world sheet itself. As a result, the Lorentz generators
transform $n_{\mu}$ and the world sheet simultaneously,
without reparametrizations.
The same property holds if one relates light-like axis $n_{\mu}$ 
with the world sheet, thus constructing Lorentz-invariant 
light cone gauge. In the paper \cite{zone} 
the $(M_{\mu\nu},\xi_{i})$ set 
was constructed with the aid of Rohrlich's parametrization, 
while in the present paper we will use for this purpose
the Lorentz-invariant light cone gauge. Below we present
a general formalism used to construct various gauges
of this kind and will select a particular one in the next Section.
We will see also that the mechanism of anomaly cancellation
from the Lorentz group is actually the same as in \cite{zone},
namely -- transition to such canonical basis, that the Lorentz
group generators $M_{\mu\nu}$ will become simple functions
of independent variables.

\begin{figure}\label{gen-scheme}
\begin{center}
~\epsfxsize=3cm\epsfysize=2.25cm\epsffile{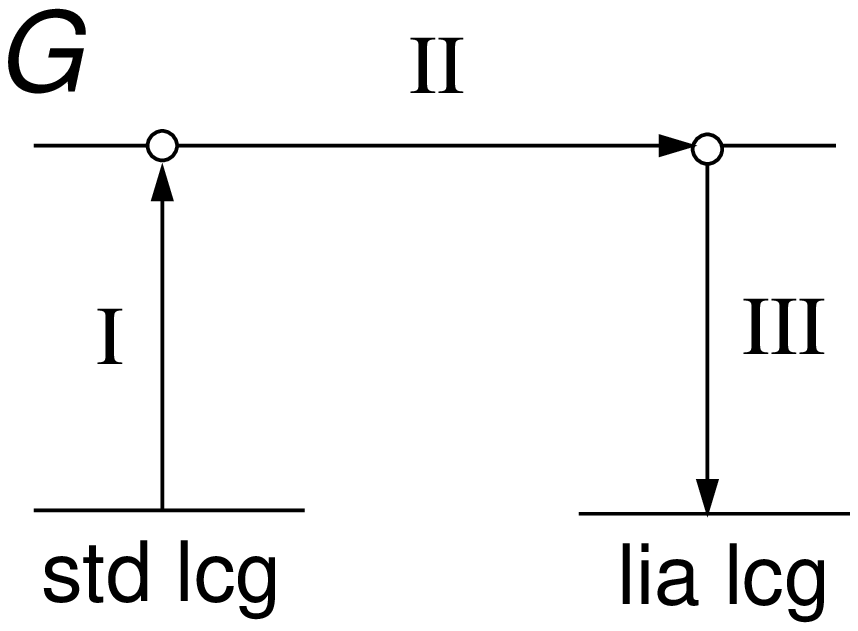}

\vspace{2mm}
\fignum Lorentz-invariant fixation of the light cone gauge.
\end{center}
\end{figure}

The construction of Lorentz-invariant light cone gauge
uses the following general scheme, see \fref{gen-scheme}.
We consider a group $G$ of gauge transformations of the
world sheet, induced by different choices of the gauge axis
$n_{\mu}$ on the light cone. It is a finite-dimensional subgroup 
in a group of general reparametrizations of the world sheet.
Standard fixation of the light cone gauge corresponds to 
a selection of one point in $G$, performed in Lorentz non-invariant way.
The first step of our construction is a recovery of $G$-symmetry in
the classical mechanics. The gauge axis $n_{\mu}$ becomes 
a dynamical variable, and the gauge group $G$ is canonically implemented
by means of explicitly constructed Hamiltonian generators.
Already at this step the Lorentz group generators $M_{\mu\nu}$
are constructed, which are gauge equivalent to the standard ones
and transform the world sheet together with the gauge axis, 
rigidly attached to it. It is shown by
explicit computation, that on quantum level the algebra of $M_{\mu\nu}$ 
has no anomalies. At this step the quantum anomaly is moved to 
the gauge group $G$.
The second step consists in a choice of a new representative
on the gauge orbit of $G$, this time related with the world sheet
in Lorentz invariant way. We select a special representative,
using gauge fixing conditions of Lorentz-invariant Abelian type, 
which lead to algebraically and geometrically simple mechanics.
At the third step the gauge is finally fixed to this representative,
and the anomalous gauge degree of freedom is eliminated.

\vspace{1mm}

Implementation of this program includes the following components.

\vspace{-1mm}
\paragraph*{Geometrical description 
of the world sheet \cite{zone,exotic}.}~
Let's introduce a function, related with string's coordinates 
and momenta by expressions

\vspace{-5mm}
\begin{eqnarray}
&&Q_{\mu}(\s)=x_{\mu}(\s)+\int_{0}^{\s}d\tilde\s \;
p_{\mu}(\tilde\s),\nn\\
&&x_{\mu}(\s)=(Q_{\mu}(\s)+Q_{\mu}(-\s))/2,\label{Qxp}\\
&&p_{\mu}(\s)=(Q'_{\mu}(\s)+Q'_{\mu}(-\s))/2\nn
\end{eqnarray}
($x,p$ are {\it even} functions of $\s$).
In terms of oscillator variables, commonly used in string theory:
\begin{eqnarray}
&&Q_{\mu}(\s)=X_{\mu}+\frac{P_{\mu}}{\pi}\sigma
+{\textstyle{{1}\over{\sqrt{\pi}}}}
\sum\limits_{n\neq0}{\textstyle{{a_{\mu}^{n}}\over{in}}}e^{in\sigma}.
\nn
\end{eqnarray}

The curve, defined by the function $Q_{\mu}(\s)$ (further called
{\it supporting curve}) has the following properties:
(i)~the curve is light-like: $Q'^{2}(\s)=0$, 
this property is equivalent to Virasoro constraints on oscillator variables;
(ii)~the curve is periodical: $Q(\s+2\pi)-Q(\s)=2P$;
(iii)~the curve coincides with the world line of one string end:
$x(0,\tau)=Q(\tau)$; the world line of another end
is the same curve, shifted onto the semi-period:
$x(\pi,\tau)=Q(\pi+\tau)-P$;
(iv)~the whole world sheet is reconstructed by this curve
as follows: $x(\s,\tau)=(Q(\s_{1})+Q(\s_{2}))/2$, $\s_{1,2}=\tau\pm\s$;
(v)~Poisson brackets for the function $Q_{\mu}(\s)$ are:
\begin{eqnarray}
&&~\hspace{-1cm}
\{Q_{\mu}(\s),Q_{\nu}(\tilde\s)\}=-2g_{\mu\nu}\vartheta(\s-\tilde\s);
\label{orig}
\end{eqnarray}
(vi)~Virasoro constraints generate reparametrizations
of supporting curve: $H=\int d\s F(\s)Q'^{2}(\s)/4$,
$\dot Q_{\mu}(\s)=\{Q_{\mu}(\s),H\}=F(\s)Q'_{\mu}(\s)$. 
Hamiltonian with $F=1$ generates uniform shifts of the argument 
$Q(\sigma)\to Q(\tau+\sigma)$, corresponding to the evolution
of the string in conformal parametrization on the world sheet:
$(\dot x\pm x')^{2}=0$.
Here $\vartheta(\s)=[\s /2\pi]+\half$, $[x]$ is integer part of $x$,
the derivative $\vartheta(\s)'=\Delta(\s)$ is periodical delta-function;
$P_{\mu}$ is total momentum of the string.

\vspace{2mm}\noindent
These properties are proven in \cite{zone} and by the other method
in~\cite{exotic}. 

\vsp\paragraph*{Mechanics in center-of-mass frame \cite{zone}.}
Let's introduce orthonormal tetrad of vectors: $N^{\alpha}_{\mu}$,
where $N^{\alpha}_{\mu}N^{\beta}_{\mu}=g^{\alpha\beta}$ and
$N^{0}_{\mu}=P_{\mu}/\sqrt{P^{2}}$. Let's decompose the supporting curve 
by this tetrad: $Q_{\mu}(\s)=N^{\alpha}_{\mu}Q^{\alpha}(\s).$

\vspace{2mm}\noindent
{\it Remark}: following to \cite{zone}, it is assumed
that variables $N^{\alpha}_{\mu}$ depend only on total momentum 
$P_{\mu}$ and constant non-dynamical vectors. Particularly, 
the space-like components $N^{i}_{\mu}$ can be explicitly constructed 
in terms of time-like vector $P_{\mu}$ and constant linearly independent 
space-like vectors $V^{i}_{\mu}$ by application of Gram-Schmidt 
orthonormalization procedure to $(P_{\mu},V^{i}_{\mu})$.
After that, the dynamical action of Lorentz group cannot 
transform $N^{\alpha}_{\mu}$ as Lorentz vectors,
but creates more complicated transformation law, see \cite{slstring}.
On the other hand, the performed substitution obviously
does not change the Poisson brackets 
$\{M_{\mu\nu},M_{\rho\s}\}$, $\{Q_{\mu},M_{\rho\s}\}$,
provided that Poisson brackets of new canonical variables $Q^{\alpha}$
are derived from old ones $Q_{\mu}$ using standard formalism, see Appendix~1.
Therefore, the variables $Q_{\mu}$ and $M_{\mu\nu}$, being expressed 
in terms of new canonical variables $Q^{\alpha}$,
are still transformed as Lorentz tensors. Further details 
of this technique are available in \cite{zone,slstring}.

\vspace{2mm}
The projection of supporting curve to CMF,
given by space-like components $Q^{i}(\s)$, is a closed curve.
The total length of this curve equals $2\sqrt{P^{2}}$.
The total area vector for the oriented surface pulled on this curve
is independent on the surface and equals $2\vec S$,
where $\vec S$ is orbital moment\footnote{
In this paper we consider bosonic string theory
and do not introduce extra spin entities.
The terms ``orbital moment'', ``generator of rotations'' 
and ``spin'' refer to the same variable, which finally determines 
the spin of elementary particles, resulting from quantization.
We use the system of units $2\pi\alpha'=c=\hbar=1$, 
where $\alpha'$ is string's tension constant, $c$ is light velocity,
$\hbar$ is Planck's constant; the metric in Minkowski space has
a signature $({+}{-}{-}{-})$.} of the string in CMF. 
These properties are also proven in \cite{zone,exotic}. 

\vsp\paragraph*{Lorentz-invariant light cone gauge \cite{ax}.}~
Virasoro constraints generate re\-pa\-ra\-me\-tri\-za\-ti\-ons 
of supporting curve, and the gauge fixing conditions 
to Virasoro constraints select particular parametrization on this curve. 
In principle, any parametrization can be constructed, 
introducing any necessary dynamical variables, but only some
of them lead to simple Hamiltonian mechanics.
Usually the following parametrization is selected:
$\s=\pi(nQ)/(nP)+Const$, i.e. for the parameter $\s$ a component 
of $Q_{\mu}$ in the direction $n_{\mu}$ is used. The vector
$n_{\mu}$ is called {\it gauge axis}. As we have mentioned above,
the choice of the gauge axis in the direction of total momentum:
$n_{\mu}=P_{\mu}/\sqrt{P^{2}}$ corresponds to time-like Rohrlich's gauge 
\cite{Rohrlich}, while the choice of the gauge axis in light-like direction: $n^{2}=0$
is called light cone gauge. In the last case the common procedure is to 
consider light-like vectors  
$n_{\pm}=(1,\pm\vec n)/\sqrt{2}$, $\vec n^{2}=1$, $n_{\pm}^{2}=0$,
$n_{+}n_{-}=1$, define the components of $Q_{\mu}'$ along $n_{\pm}$ as
$Q_{\pm}'=(n_{\mp}Q')$, and the space-like component,
orthogonal to $\vec n$, as $\vec Q_{\perp}'$, 
select the gauge axis in the direction $n_{-}$: $Q_{+}'=P_{+}/\pi$, 
and using the condition ${Q'}^{2}=2Q_{+}'Q_{-}'-\vec Q_{\perp}'{}^{2}=0$,
finally express the remaining component in terms of $P$ and $\vec Q_{\perp}'$:
$Q_{-}'=\pi({\vec Q}_{\perp}')^{2}/2P_{+}$.
Then, substituting Fourier expansion for $\vec Q_{\perp}'$, we can 
express the whole mechanics in terms of total momentum, mean coordinate
and Fourier coefficients (transverse oscillator variables).

Let's introduce the following parametrization on the supporting curve:
\begin{eqnarray}
&&~\hspace{-8mm}Q^{\alpha}(\sigma) = Q^{\alpha}(0)+\int_{0}^{\sigma}
     d\sigma ' a^{\alpha}(\sigma '),\label{Qdlcg}\\
&&~\hspace{-8mm}a^{0}(\sigma) =  
 \frac{\pi}{2\sqp}\left(\frac{P^{2}}
  {\pi^{2}}+|a(\sigma)|^{2}\right),\nn\\
&&~\hspace{-8mm}\vec a(\sigma) =  
\frac{a(\sigma)+\astar(\sigma)}{2} \; {\vec e}_{1} +
\frac{a(\sigma)-\astar(\sigma)}{2} \; i{\vec e}_{2}+\nn\\
&&+\frac{\pi}{2\sqp}\left(\frac{P^{2}}{\pi^{2}}-|a(\sigma)|^{2}
\right){\vec e}_{3},\nn
\end{eqnarray}
where $a(\sigma)=\sqrt{\frac{2}{\pi}}\sum\limits_{n\neq0} a_{n}e^{-in\sigma}$
and ${\vec e}_{k}$ is an orthonormal basis in CMF.
Here we easily recognize the light cone gauge
$Q^{+}(\sigma)\equiv \nm\q=Q^{+}(0)+\frac{1}{\pi}
     \sqrt{\frac {P^{2}}{2}}\;\sigma$ with the gauge axis
$\nm=\frac{1}{\sqrt{2}}(N_{\mu}^{0}-N_{\mu}^{i}e_{3}^{i})$.
The difference from standard approach is that $\nm$ is now {\it dynamical}
variable, because we do not fix ${\vec e}_{3}$ to a constant vector.
Using this parametrization, we obtain the following mechanics:

\begin{center}
$P_{\mu},Z_{\mu}$ +  infinite set  of  oscillators $a_{k},a_{k}^{*}$ +\\
+  the top ${\vec e_{i}},{\vec S}$,
\end{center}

\noindent
with Poisson brackets, derived in Appendix~1:
\begin{eqnarray}
&&\{Z_{\mu},P_{\nu}\}=g_{\mu\nu},  \nn\\
&&\{a_{k},a_{n}^{*}\}=ik\delta_{kn},\ k,n\in\Z\backslash\{0\},\label{Pb1}\\
&&\{S^{i},S^{j}\}=-\epsilon^{ijk}S^{k},  \
    \{S^{i},e^{j}_{n}\}=-\epsilon^{ijk}e^{k}_{n}.\nn
\end{eqnarray}
Here $Z_{\mu}= {\textstyle{{1}\over{2\sqrt{P^{2}}}}}\int_{0}^{2\pi}d\s\; 
a^{0}(\sigma)(Q_{\mu}(\sigma)-({\textstyle{{\sigma}\over{\pi}}}
-1)P_{\mu})+\half \epsilon^{ijk}\Gamma _{\mu}^{ij}S^{k}$
is Newton-Wigner mean coordinate \cite{zone}, conjugated to $P_{\mu}$; 
$\kr$ are Christoffel symbols
and $\sbo=-\frac{1}{4}\int_{0}^{2\pi}d\sigma\;\int_{0}^{\sigma} d\sigma '\;
    {\vec a}(\sigma)\times {\vec a}(\sigma ')$ 
is an orbital moment of the string in CMF.

The mechanics is restricted by four constraints of the 1st class:
$$\{\chi_{0},\chi_{i}\}=0\ ,\ 
\{\chi_{i},\chi_{j}\}=\epsilon_{ijk}\chi_{k},$$
which include mass shell condition and requirements
of the form ``spin of the top is equal to the spin of the string'': 
\begin{eqnarray}
&& \chi_{0}={\textstyle\frac{P^{2}}{2\pi}}-L_{0}=0,\quad 
L_{0}=\sum\limits_{n\neq0} a_{n}^{*}a_{n},\nn\\
&&\chi_{3}=S_{3}-A_{3}=0,\quad  
A_{3}=\sum\limits_{n\neq0} {\textstyle\frac{1}{n}}a_{n}^{*}a_{n},\nn\\ 
&&\chi_{+}= S_{+}-A_{+}=0,\quad  \chi_{-}=S_{-}-A_{-}=0,\nn\\
&&\chi_{\pm}=\chi_{1}\pm i\chi_{2},\quad  S_{\pm}=S_{1}\pm iS_{2},\nn\\
&& A_{-}=\sqrt{{\textstyle\frac{2\pi}{P^{2}}}}\sum_{k,n,k+n\neq0} 
{\textstyle\frac{1}{k}}a_{k}a_{n}a_{k+n}^{*},\label{Apol}\\
&& A_{+}=\sqrt{{\textstyle\frac{2\pi}{P^{2}}}}\sum_{k,n,k+n\neq0} 
{\textstyle\frac{1}{k}}a_{k}^{*}a_{n}^{*}a_{k+n},\nn
\end{eqnarray}
where $S_{i}=S^{k}e_{i}^{k}$ is a projection of ${\vec S}$ onto 
${\vec e}_{i}$. According to Dirac's theory of constrained
mechanical systems \cite{Dirac}, the constraints of the 1st class
generate gauge transformations:
\begin{itemize}
\item $\chi_{0}$ generates phase shifts of oscillator variables
$E_{0}: a_{n}\to a_{n}e^{-in\tau}$ and translations of mean 
coordinate $Z\to Z+P\tau/\pi$; these transformations
shift the argument $Q(\s)\to Q(\s+\tau)$; equivalently
they produce a reparametrization of the world sheet,
related with the evolution of the string;
\item $\chi_{3}$ generates phase shifts $R_{3}: a_{n}\to a_{n}e^{-i\alpha}$
and rotations of $\vec e_{1,2}$ about $\vec e_{3}$:
$\vec e_{i}\to R(\vec e_{3},-\alpha)\vec e_{i},\ i=1,2$;
these transformations preserve $Q(\s)$ and points on the world sheet;
\item $\chi_{1,2}$ generate rotations of basis $\vec e_{i}$
about axes $\vec e_{1,2}$ and certain non-linear transformations
of oscillator variables; these transformations change the direction
of the gauge axis and perform corresponding reparametrizations 
of the supporting curve and the world sheet.
\end{itemize}

The obtained parametrization of the world sheet
differs from the standard light cone gauge by the introduction 
of six variables $(\vec e_{i},\vec S)$, forming the mechanics of the top.
It also includes three constraints of the 1st class, which directly eliminate
three degrees of freedom ($\vec S=\vec A$), and generate gauge transformations,
identifying three remaining ones (physical observables are independent on
the choice of $\vec e_{i}$). According to Dirac's theory of constraints \cite{Dirac}, 
such implementation leads to equivalent mechanical system.
At this point we have recovered the gauge symmetry $G$, 
mentioned in Section~\ref{gen-concepts}, in the classical mechanics.
It is the group of the world sheet reparametrizations, related with 
a different choices of the gauge axis. Now
the gauge axis is a dynamical variable, presented
by the unit norm 3-vector $\vec e_{3}$ in the center-of-mass frame,
and by the above defined light-like 4-vector $n_{\mu}^{-}$ in Minkowski space.
$G$-symmetry is generated canonically by the constraints $\chi_{i}$,
with factorization by the trivial subgroup $R_{3}$, and possesses  
the group manifold $G=SO(3)/SO(2)=S^{2}$.

\vsp\paragraph*{Lorentz group generators \cite{zone}}  
are given by expression 
\begin{eqnarray}
&&M_{\mu\nu}=\int_{0}^{\pi}d\s(x_{\mu}p_{\nu}-x_{\nu}p_{\mu})=\nn\\
&&=X_{\mu}P_{\nu}-X_{\nu}P_{\mu}
+\epsilon_{ijk}N_{\mu}^{i}N_{\nu}^{j}S^{k},\nn\\
&&X_{\mu}=Z_{\mu}-\half \epsilon_{ijk}\Gamma _{\mu}^{ij}S^{k},\label{Xdef}
\end{eqnarray}
they generate Lorentz transformations of a coordinate frame
$(N_{\mu}^{0},N_{\mu}^{k}e_{i}^{k})$, by which the configuration
is decomposed with scalar coefficients. Thus, $M_{\mu\nu}$ generate
``rigid'' Lorentz transformations of the world sheet,
not changing its parametrization. Lorentz generators are
in involution with constraints: $\{M_{\mu\nu},\chi_{0,i}\}=0$.
This fact means simultaneously that the generators of gauge 
transformations $\chi_{0,i}$ are Lorentz-invariant 
and generators of Lorentz group $M_{\mu\nu}$ are gauge-invariant.

Lorentz generators are simple functions (\ref{Xdef})
of variables $(Z,P,\vec S)$,
which in our approach are independent, i.e. their quantum commutators
are postulated directly from Poisson brackets.
In \cite{slstring} it has been shown by direct calculation
that under these conditions the quantum commutators
$[M_{\mu\nu},M_{\rho\s}]$ are anomaly free. 
We present this computation in Appendix~2.
This result is not surprising, because after the performed formal
substitutions the generators of rotations in CMF became
independent canonical variables. At this point the anomaly 
is not removed from the theory, but {\it transferred} 
from the Lorentz group to the gauge group $G$.
Indeed, the constraints $\chi_{i}$ are
cubic in terms of oscillator variables and in quantization their algebra 
acquires exactly the same anomaly that earlier was in Lorentz group.
Now we should do the second step of the diagram \fref{gen-scheme}: 
select the gauge fixing conditions, which will eliminate 
the gauge freedom associated with $G$-symmetry. 
The gauge fixing conditions of the type $\vec e_{3}=(1,0,0)$
would return us immediately to the standard light cone gauge,
with the anomaly in Lorentz group. We will use the alternative
gauge fixing conditions, which relate $\vec e_{3}$ 
with respect to other dynamical vectors in the system,
and introduce Lorentz-invariant parametrization on the world sheet.

At first, we will represent the straight-line string solution \cite{slstring} 
in the constructed light cone gauge. For this solution the projection
of supporting curve to CMF is a circle, and the string itself
has a form of a straight segment rotating about its middle 
at constant angular velocity. According to \cite{slstring},
this solution allows anomaly-free quantization at arbitrary 
dimension $d\geq3$. It belongs to a border of classical Regge-plot 
$P^{2}/2\pi\geq S$, and in quantum theory corresponds 
to a leading Regge-trajectory $P^{2}/2\pi=S$. In our system
of variables this solution corresponds to 
a single excited mode $a_{1}$, if the gauge axis $\vec e_{3}$ 
is directed along the spin $\vec S$:
\begin{eqnarray}
&&~\hspace{-15mm}
a_{n}=0,\ n\neq1,\ S_{\pm}=0,\
{\textstyle\frac{P^{2}}{2\pi}}=S=S_{3}=|a_{1}|^{2}.\label{1mode}
\end{eqnarray}
Further we will refer to this case as northern pole solution 
(where the circle $\vec Q(\s)$ defines the equator).
There is a gauge-equivalent one-modal solution with exited $(a_{-1})$-mode
and $\vec e_{3}$ opposite to $\vec S$ (southern pole solution). 
Other directions of gauge axis for the straight-line string 
give infinitely-modal solutions. 
Later we will use the solution (\ref{1mode}) to study the structure 
of general theory in its vicinity. 

\vsp\paragraph*{Gauge fixing conditions} we propose 
have a form
\begin{eqnarray}
&&a_{s}+a_{-s}=0,\quad a_{s}^{*}+a_{-s}^{*}=0\label{gauge}
\end{eqnarray}
for some $s>0$. The straight-line 
string solution (\ref{1mode}) satisfies these conditions
at $s>1$. Later, in Section~\ref{gribsec},
we will show that conditions (\ref{gauge})
can be imposed on any solution of string theory.

The gauge fixing conditions (\ref{gauge}) are preserved by
transformations $R_{3}$. They are not preserved by $E_{0}$,
however, there is a remainder of $E_{0}$-symmetry,
discrete transformation 
$$D_{2s}:\ a_{n}\to a_{n}e^{-in\pi/s}$$
preserving (\ref{gauge}): $a_{s}+a_{-s}=0\to-a_{s}-a_{-s}=0$.
Therefore, $R_{3},D_{2s}$-symmetries are present in the theory 
after gauge fixation.

The procedure of gauge fixation, described in Appendix~1,
is the third step on the diagram \fref{gen-scheme}. 
It results to the same canonical basis as (\ref{Pb1}),
but without $a_{\pm s}$ oscillators. This exclusive property
(simplicity of Poisson brackets)
follows from the fact that two gauge fixing conditions (\ref{gauge})
are in involution with each other: $\{a_{s}+a_{-s},a_{s}^{*}+a_{-s}^{*}\}=0$,
i.e. generate {\it Abelian} group of transformations.
By this fact our approach differs from non-Abelian gauges, 
considered in \cite{partI-III} Part~I, 
which possess complicated Poisson brackets.
Further we refer to the constructed parametrization
as Lorentz-invariant Abelian light cone gauge \hbox{(lia-lcg)}.

The oscillators $a_{\pm s}$ now become dependent variables,
whose expressions should be found from $\chi$-constraints.
The contribution of $a_{\pm s}$ in $A_{3}$ 
vanishes: ${\textstyle{1\over s}}(|a_{s}|^{2}-|a_{-s}|^{2})=0$,
as a result, $a_{\pm s}$-terms drop out from $\chi_{3}=S_{3}-A_{3}$.
This result follows from the fact that gauge fixing conditions (\ref{gauge}) 
are preserved by transformation $R_{3}$ and therefore are 
in involution with $\chi_{3}$.
The gauge fixing conditions (\ref{gauge}) 
are not in involution with $\chi_{0}$, 
and $a_{\pm s}$-terms in $L_{0}$ do not vanish: 
$|a_{s}|^{2}+|a_{-s}|^{2}=2|a_{s}|^{2}$. Poisson brackets of (\ref{gauge})
with $\chi_{\pm}$ also do not vanish. We conclude that (\ref{gauge}) 
are gauges for $(\chi_{\pm},\chi_{0})$, and $a_{\pm s}$ should be 
determined from these three constraints.

\vsp\section{Algebraic properties of lia-lcg}
\label{algsec}
Isolating contribution of $a_{\pm s}$-oscillators in (\ref{Apol})
and using the relation $a_{-s}=-a_{s}$, we have
\begin{eqnarray}
&&~\hspace{-8mm}
a_{s}^{2}d+\half a_{s}a_{s}^{*}d^{*}+a_{s}f+a_{s}^{*}g
+\Sigma_{-}-\sqrt{{\textstyle{P^{2}\over2\pi}}}S_{-}=0,\nn\\
&&~\hspace{-8mm}
a_{s}^{*2}d^{*}+\half a_{s}a_{s}^{*}d+a_{s}^{*}f^{*}+a_{s}g^{*}
+\Sigma_{+}-\sqrt{{\textstyle{P^{2}\over2\pi}}}S_{+}=0,\nn\\
&&~\hspace{-8mm}
{\textstyle{P^{2}\over2\pi}}=L_{0}^{(s)}+2a_{s}a_{s}^{*},\label{maineqs}
\end{eqnarray}
where
\begin{eqnarray}
&& d=d_{+}-d_{-},\ f=f_{+}-f_{-},\ g=g_{-}-g_{+},\nn\\
&&d_{+}=a_{2s}^{*}/s,\ d_{-}=a_{-2s}^{*}/s,\
L_{0}^{(s)}={\sum}'a_{k}^{*}a_{k}\nn\\
&&g_{-}={\sum}'
{\textstyle{1\over k}}a_{k}a_{s-k},\ 
g_{+}={\sum}'
{\textstyle{1\over k}}a_{k}a_{-s-k},\nn\\
&&f_{+}={\sum}'
\left({\textstyle{1\over s}}+{\textstyle{1\over k}}\right)
a_{k}a_{s+k}^{*},\label{defelem}\\ 
&&f_{-}={\sum}'
\left(-{\textstyle{1\over s}}+{\textstyle{1\over k}}\right)
a_{k}a_{-s+k}^{*},\nn\\
&&\Sigma_{-}={\sum}'
{\textstyle{1\over k}}a_{k}a_{n}a_{k+n}^{*},\
\Sigma_{+}={\sum}'
{\textstyle{1\over k}}a_{k}^{*}a_{n}^{*}a_{k+n},\nn\\
&&\Sigma_{-}^{*}=\Sigma_{+},\nn
\end{eqnarray}
Here in the sums ${\sum}'$ terms with $a_{0}^{(*)}$ and
$a_{\pm s}^{(*)}$ are excluded. Introducing denotations
\begin{eqnarray}
&&\lambda=\sqrt{{\textstyle{P^{2}\over2\pi}}},\
n_{s}=(\lambda^{2}-L_{0}^{(s)})/2,\nn\\
&&k=\Sigma_{-}+\half n_{s}d^{*}-\lambda S_{-},\nn
\end{eqnarray}
we can treat (\ref{maineqs}) as overdetermined polynomial system
for $(a_{s},a_{s}^{*})$ at given values of coefficients
$(d,f,g,k,n_{s})$:
\begin{eqnarray}
&&a_{s}^{2}d+a_{s}f+a_{s}^{*}g+k=0,\label{maineqs1}\\
&&a_{s}^{*2}d^{*}+a_{s}^{*}f^{*}+a_{s}g^{*}+k^{*}=0,\nn\\
&&a_{s}a_{s}^{*}-n_{s}=0.\nn
\end{eqnarray}

In \cite{partI-III} this system was solved analytically
using the technique of Groebner's basis \cite{ideals}.
The result comprises a polynomial condition of consistency 
for the system (\ref{maineqs1}) and rational expressions for $a_{s}$. 
The condition of consistency is the Dirac's constraint, 
remaining after imposition of two gauge fixing
conditions (\ref{gauge}) to three original constraints 
$\chi_{0},\chi_{\pm}$, it is a polynomial equation 
of 8th order in $\lambda=\sqrt{P^{2}/2\pi}$, playing a role of new
mass shell condition. Being used as Hamiltonian, it generates 
correct string evolution, consisting of shifts $\s\to\s+\tau$
and such reparametrizations that keep gauge fixing conditions 
(\ref{gauge}) permanently satisfied. Further analysis 
of this mechanics can be found in \cite{partI-III}.

For the construction of quantum theory it's more convenient
to apply a different form of the mechanics, using expansion series
in the vicinity of straight-line string (\ref{1mode}).

\vsp\paragraph*{Solutions in the vicinity of straight-line string}~

\vspace{2mm}\noindent
Computing the Jacobian of the system (\ref{maineqs}) on the
straight-line string's northern pole solution (\ref{1mode}),
we see that this system is non-degenerate at $s=2$.
In this case in the vicinity of (\ref{1mode}) it has a unique solution, 
representable as $C^{\infty}$-smooth analytical function 
of coefficients of the system. The solution is given by series:
\begin{eqnarray}
&&a_{2}=\sum\limits_{n\geq1}{P_{n}a_{1}^{2}\over|a_{1}|^{4n}},\label{a2cl}
\end{eqnarray}
where $P_{n}$ are polynomials of $a_{1},\Sigma_{-},S_{-},d,f,g'$,
their conjugates and $\gamma,\gamma^{-1}$. Here 
$g'_{-}=g_{-}-a_{1}^{2}$, $g'=g'_{-}-g_{+}$,
$\gamma=(L_{0}^{(2)})^{-1/2}$. 
Explicit expressions for the first three polynomials $P_{n}$ are
given in Appendix~3, their general properties are studied
in \cite{partI-III}.
For the purposes of further consideration it is 
convenient to extract from $a_{2}$ a common phase multiplier 
$a_{1}^{2}/|a_{1}|^{2}$ and to define
$a_{2}=\alpha_{2}\cdot a_{1}^{2}/|a_{1}|^{2}$, so that
$|a_{2}|=|\alpha_{2}|$ and

\begin{eqnarray}
&&~\hspace{-10mm}
\alpha_{2}=\sum\limits_{n\geq1}{P_{n}\over|a_{1}|^{4n-2}},\quad
{P^{2}\over2\pi}=L_{0}^{(2)}+2|\alpha_{2}|^{2}.\label{alp2cl}
\end{eqnarray}

\vsp\section{Geometric properties of lia-lcg}
\label{gribsec}
In this Section we describe in more details the transformations, 
generated by constraints $\chi_{0},\chi_{i}$. At first, we introduce
several definitions.

Let's consider in 3D space: smooth closed curve $\vec Q(\s)$ 
with marked point $O$ 
and unit norm vector $\vec e_{3}$. Let's introduce variables
\begin{eqnarray}
&&~\hspace{-10mm}
{\vec a}_{n}={\textstyle{{1}\over{2\sqrt{2\pi}}}}
\oint d\vec Q(\s)\cdot\label{oint}\\
&&\cdot\exp\left[
{\textstyle{{2\pi in}\over{L_{t}}}}
\left(L(\s)-(\vec Q(\s)-\vec Q(0))\vec e_{3}\right)\right],\nn
\end{eqnarray}
where $L(\s)$ is a length of arc between points $O$
and $\vec Q(\s)$ along the curve, $L_{t}$ is total length of the curve.
Two properties obviously follow from the definition:
${\vec a}_{-n}={\vec a}_{n}^{*},\ {\vec a}_{0}=0.$ 
Let's decompose vectors ${\vec a}_{n}$ into the components, parallel and
orthogonal to $\vec e_{3}$: 
${\vec a}_{n}=a_{n3}\vec e_{3}+{\vec a}_{n\perp}$,
and denote real and imaginary parts of ${\vec a}_{n\perp}$
as ${\vec q}_{n\perp}$ and ${\vec p}_{n\perp}$: 
${\vec a}_{n\perp}={\vec q}_{n\perp}+i{\vec p}_{n\perp}$. 
Let's fix some $n=s>0$ and write ${\vec q}_{s\perp}={\vec q}$ 
and ${\vec p}_{s\perp}={\vec p}$.
Functions ${\vec q}(\vec e_{3}),{\vec p}(\vec e_{3})$ 
define smooth vector fields on unit sphere of $\vec e_{3}$
(tangent to the sphere). Due to topological ``hedgehog'' theorem, 
these fields have singular points on the sphere, 
where corresponding field vanishes, e.g. ${\vec q}=0$.

\vspace{2mm}
\noindent{\it Remark:} 
the curve $\vec Q$ is a projection of supporting curve to CMF.
Gauge axis $\vec e_{3}$ relates 
the following parametrization to this curve:
\begin{eqnarray}
&&\s={{{2\pi}\over{L_{t}}}}(L(\s)-Q_{3}(\s)+Q_{3}(0)),
\label{lcgdef}
\end{eqnarray}
where $Q_{3}=\vec Q\vec e_{3}$. Now we recognize in (\ref{oint}) 
Fourier modes of function $\vec a(\s)=\vec Q'(\s)$, 
where $\s$ is lcg-parameter (\ref{lcgdef}). This expression
is written in parametric-invariant form, as circulation integral.
Such form of definition is also known in string theory
as DDF variables \cite{DDF}.
Vectors $\vec a_{n\perp}$ are related with earlier introduced 
oscillator variables $a_{n}$ as follows:
\begin{eqnarray}
&&\vec a_{n\perp}=a_{n1}\vec e_{1}+a_{n2}\vec e_{2},\nn\\
&&a_{n1}=(a_{n}+a_{-n}^{*})/2,\ a_{n2}=i(a_{n}-a_{-n}^{*})/2.\nn
\end{eqnarray}

\vspace{-5mm}
\begin{figure}\label{qpfields}
\begin{center}
~\epsfxsize=5cm\epsfysize=2.3cm\epsffile{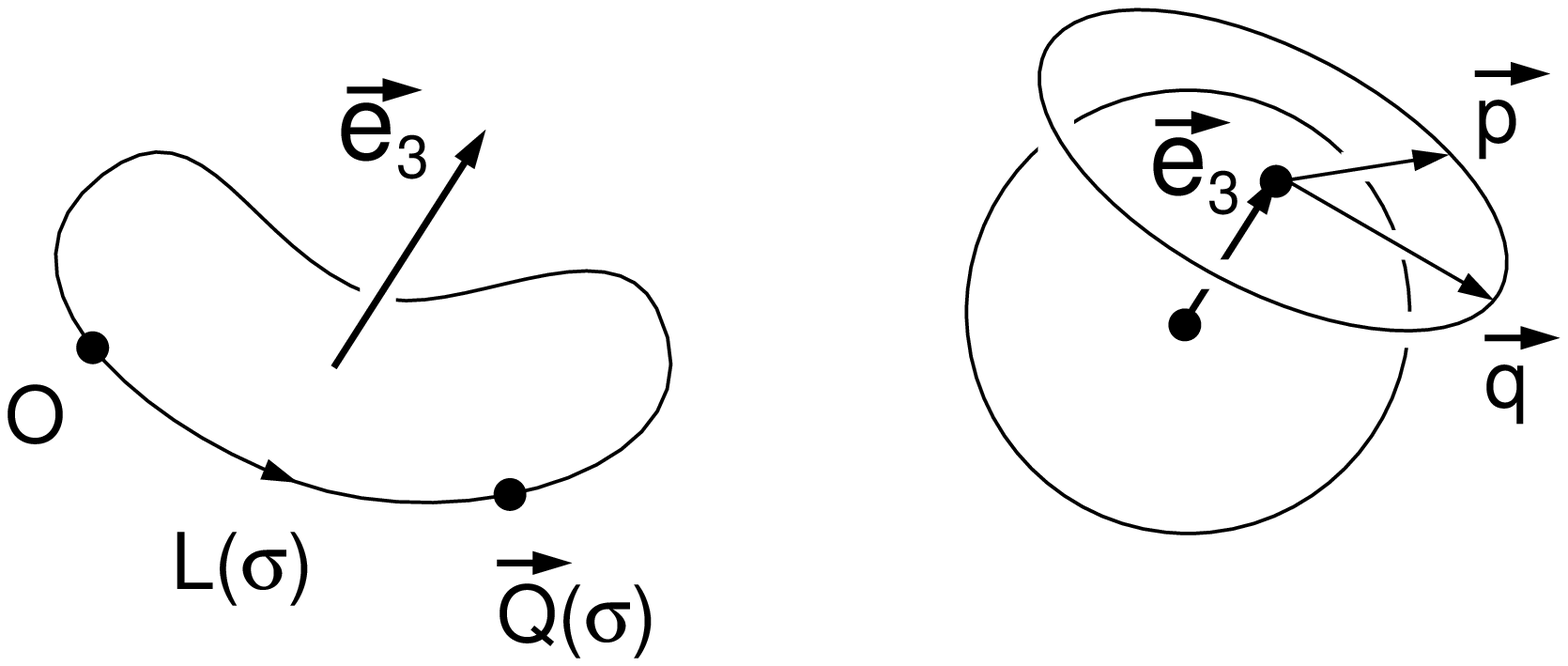}

\fignum Definition of vector fields $\vec q(\vec e_{3}),\vec p(\vec e_{3})$.

\end{center}
\end{figure}

From the above definitions the following properties become clear:
$\chi_{3}$ generates transformations, preserving
$\vec e_{3}$ and $\vec a_{n\perp}$; $\chi_{1,2}$ generate rotations of
$\vec e_{3}$ and associated changes of $\vec a_{n\perp}$
according to the formula (\ref{oint}). 
Gauge fixing conditions (\ref{gauge}) correspond 
to a singular point ${\vec q}=0$. 
Evolution, generated by $\chi_{0}$, is represented as phase rotations 
${\vec a}_{n\perp}\to{\vec a}_{n\perp}e^{-in\tau}$, equivalent to
a motion of vectors $\vec q$ and $\vec p$ along the ellipse,
shown at \fref{qpfields} on the right. During the evolution 
a vector $\vec\omega=\vec q\times\vec p$ is preserved,
and singular points ${\vec q}=0$ move along zero-level curves
of a function $F(\vec e_{3})=\vec\omega\vec e_{3}=0$.
On these curves the ellipse shown at \fref{qpfields}-right
degenerates to a segment.

\vspace{2mm}
From here we see that gauge fixing conditions (\ref{gauge})
can be imposed on any solution of string theory.
Namely, these conditions can be satisfied for any curve 
$\vec Q(\s)$, directing the light cone gauge axis $\vec e_{3}$
to a singular point $\vec q(\vec e_{3})=0$ 
of a vector field on the sphere, constructed in terms
of this curve. Therefore, $\vec e_{3}$ is implicitly expressed 
in terms of $\vec Q(\s)$, which is uniquely related with the variables 
of original description: coordinates and momenta $x(\s),p(\s)$. 
The vectors $\vec e_{1,2}$ are not fixed and can freely rotate
about $\vec e_{3}$. This gauge transformation is generated by
the constraint $\chi_{3}$, which can be preserved in the theory,
because its non-linearity is insufficient to create any anomalies.

Now let's return to the consideration of singular points 
$\vec q(\vec e_{3})=0$. In general position the vector fields 
on the sphere have even number of singular points, which is $\geq2$. 
We remind that non-degenerate singular points of 2-dimensional vector 
fields are~\cite{DNF}:

\begin{figure}\label{singtypes}
\begin{center}
~\epsfysize=1.5cm\epsfxsize=6cm\epsffile{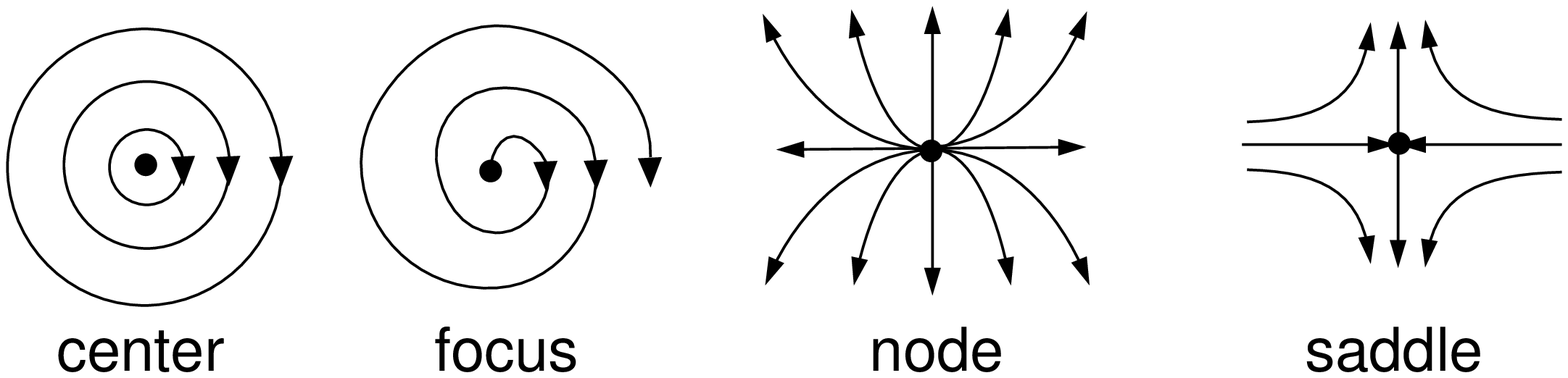}

\fignum Non-degenerate singular points\\ of 2-dimensional vector field.

\end{center}
\end{figure}

\noindent{\it Definition 1:} For each type 
{\it index of singularity} is defined as algebraic number of rotations
of vector $\vec q(\vec e_{3})$, when $\vec e_{3}$ passes around singular point
($>0$, if directions of rotations of $\vec q$ and $\vec e_{3}$ coincide;
$<0$ otherwise). For (center, node, focus) it is $+1$, for saddle $-1$.

\vspace{2mm}
The sum of all indices is equal to Euler characteristic of the surface,
defined as (num. of vertices) $-$ (num. of edges) $+$ (num. of faces)
for any tessellation of the surface. For the sphere this characteristic
equals 2, thus generic vector field on a sphere has
2 singular points of index $+1$ and arbitrary number 
of self-compensating pairs $(+1,-1)$. In special cases 
there can be degenerate singularities of more complex form,
e.g. multi-saddles obtained by a fusion of several saddles.

Presence of several singular points $\vec q(\vec e_{3})=0$ 
leads to the fact that an orbit of the gauge group generated 
by $\chi$-constraints intersects the surface of gauge fixing conditions 
(\ref{gauge}) in several points of the phase space. 
Because these points are gauge equivalent, the mechanics possesses
{\it discrete} gauge symmetry. This phenomenon encountered 
in the theory of non-Abelian gauge fields, where it has been studied by
V.N.Gribov~\cite{Gribov}. In full generality this question was addressed 
in the paper \cite{nogauge}. We will call such equivalent points 
{\it Gribov's copies}. 

The following topological invariant can be introduced 
for Gribov's copies.  

\begin{figure}\label{orbit}
\begin{center}
~\epsfxsize=4cm\epsfysize=2.5cm\epsffile{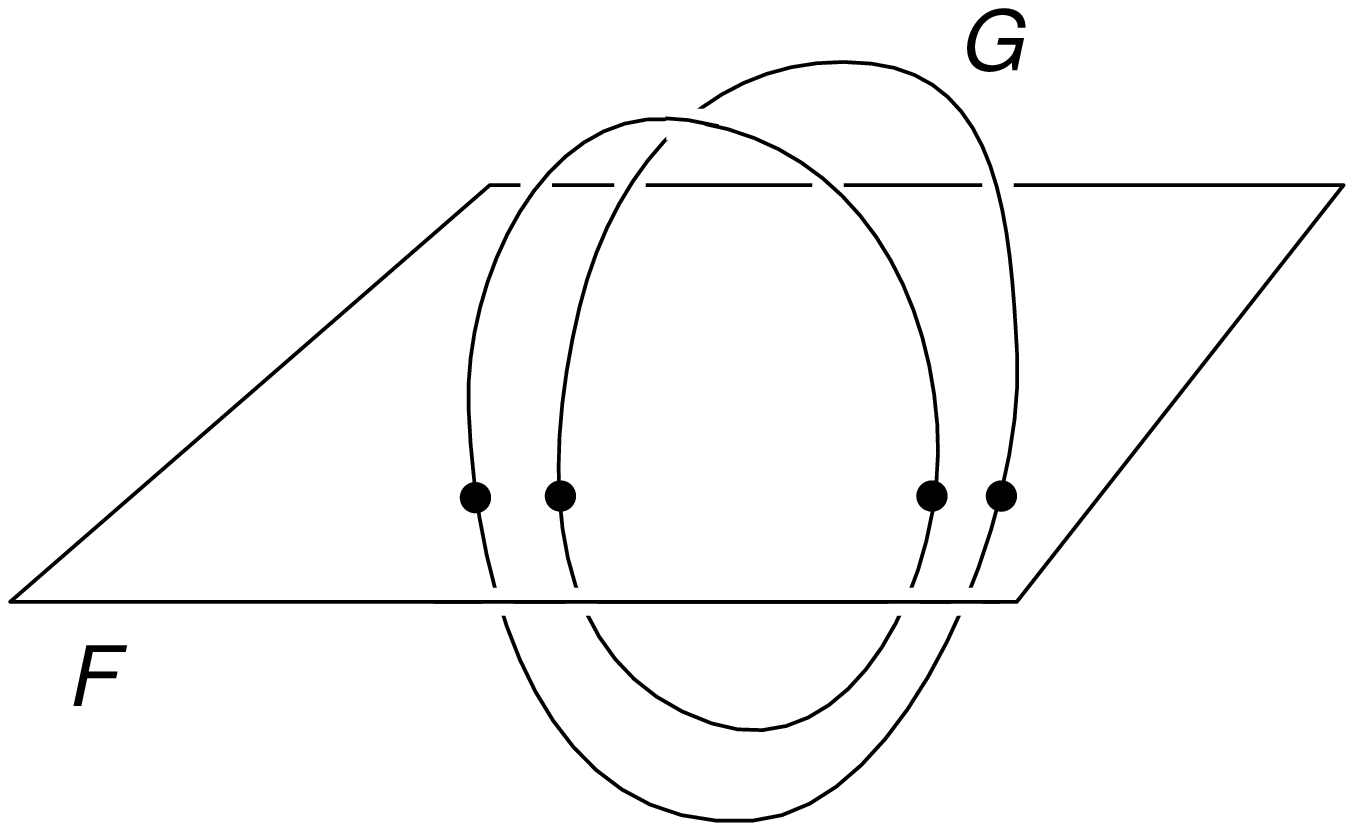}

\end{center}

\vspace{-3mm}
\fignum Gribov's copies: the orbit of a gauge group $G$
intersects the surface of gauge fixing condition $F$
in several points.

\end{figure}

\vspace{2mm}
\noindent{\it Definition 2:} let the phase space $M$ be 
a smooth orientable manifold. Let the orbit of gauge group $G$ 
and the surface of gauge fixing condition $F$ be its smooth 
orientable submanifolds with $dim(G)=codim(F)$. 
Let $P$ be the point of their transversal intersection.
This means that in point $P$ the tangent spaces to $G$ and $F$ 
span the tangent space to $M$. Let $\vec\tau(F,P)$,
$\vec\tau(G,P)$ and $\vec\tau(M,P)$ be the bases 
in the tangent spaces, defining the orientation of $F$, $G$ and $M$,
evaluated in point $P$. {\it The index of intersection} 
of $F$ and $G$ in point $P$ is defined as a number $\nu$,
equal to $(+1)$ if the basis $(\vec\tau(F,P),\vec\tau(G,P))$ 
has the same orientation as the basis $\vec\tau(M,P)$,
and equal to $(-1)$ if the orientations are opposite.

\vspace{2mm}
It has been shown in \cite{partI-III} that
for the Gribov's copies of string theory 
the definitions 1 and 2 coincide.

\vspace{2mm}
In lia-lcg the copies comprise different sets of oscillator variables 
$\{a_{n}\}$, which reproduce the same but differently parameterized 
curves $\vec Q(\s)$. As a result, Gribov's copies correspond to
discrete reparametrizations of the world sheet.
To identify the equivalent states, in classical theory 
the phase space should be factorized with respect to this symmetry. 
Analogous procedures can be applied in quantum mechanics, 
e.g. by constructing irreducible representations for operators 
possessing this discrete symmetry and formulating respective selection rules.
However, on the quantum level the discrete non-linear reparametrizations 
of the world sheet can be violated by anomalies, excluding this
symmetry from the theory.

The explicit formulae for the vector fields 
are given by (\ref{oint}). There is also an alternative definition:
\begin{eqnarray}
&&~\hspace{-10mm}
{\vec a}_{n}={\textstyle{{1}\over{2\sqrt{2\pi}}}}
\oint d\vec Q(\s)\cdot\label{oint2}\\
&&\cdot\exp\left[
{\textstyle{{2\pi in}\over{L_{t}}}}
\left(L(\s)-(\vec Q(\s)-\vec X)\vec e_{3}\right)\right],\nn
\end{eqnarray}
where $\vec X=\oint dL(\s)\vec Q(\s)/L_{t}$ 
defines an average position of the curve $\vec Q(\s)$
and coincides with the definition of mean coordinate (\ref{Xdef})
projected to CMF. The difference of (\ref{oint}) and (\ref{oint2})
consists in $\vec e_{3}$-dependent 
phase factor $e^{in\varphi(\vec e_{3})}$, $\varphi(\vec e_{3})=
{2\pi}(\vec X-\vec Q(0))\vec e_{3}/{L_{t}}$.
Because the evolution of ${\vec a}_{n}$ 
is the phase rotation, the phase factor actually introduces 
a difference of ``local time'' for the evolution of vector fields
on the sphere. This factor preserves the orbits of 
Gribov's copies and changes their evolution parameter
from lcg's (\ref{lcgdef}) to the natural one (length of the curve).
The definition (\ref{oint2}) is
more convenient to describe the evolution of Gribov's copies.
In \cite{partI-III} the structure of Gribov's copies 
for nearly straight solution (\ref{1mode}) has been investigated
using computer visualization of the vector fields and
analytical criteria. It is shown, that
for the straight-line string $\vec Q_{0}(\s)=(\cos\s,\sin\s,0)$ 
the vector field $\vec q_{s\perp}(\vec e_{3})$ has the following
singular points: at $s=1$ two nodes on the equator;
at $s=2$ saddles on the northern and southern poles
and four nodes on the equator;
at $s>2$ multi-saddles on the northern and southern poles
and $2s$ nodes on the equator.
During the evolution the nodes move along the equator,
while the singular points in the poles stay fixed
and saddle patterns rotate around them. After a small deformation 
of the string from straight configuration $\vec Q(\s)=
\vec Q_{0}(\s)+\delta \vec Q(\s)$ the nodes
move along a common trajectory in the vicinity of the equator;
at $s=2$ the saddles move in small loops near the poles; at $s>2$ 
the multi-saddles are unfolded to $(s-1)$ non-degenerate saddles
moving near the poles. After the lapse of time 
$\Delta\tau_{s}=\pi/s=$(period of evolution)$/2s$ 
the vector field reverses its direction and the pattern of 
singularities returns to the initial state ($D_{2s}$-symmetry). 
During this time the equatorial singularities move to the neighbor ones, 
pole singularities at $s=2$ perform one revolution along the loops.

\vspace{-5mm}\section{Quantum mechanics}

\vspace{-2mm}\paragraph*{Canonical operators}~
\begin{eqnarray}
&&[Z_{\mu},P_{\nu}]=-ig_{\mu\nu},\nn\\
&&[a_{k},a_{n}^{+}]=k\delta_{kn},\ k,n\neq0,\pm s,\label{canon}\\ 
&&[S^{i},S^{j}]=i\epsilon_{ijk}S^{k},\
[S^{i},e_{j}^{k}]=i\epsilon_{ikl}e_{j}^{l},\nn\\
&&[S_{i},S_{j}]=-i\epsilon_{ijk}S_{k},\
[S_{i},e_{j}^{k}]=-i\epsilon_{ijl}e_{l}^{k},\nn\\
&&[S^{i},S_{j}]=0,\ e_{i}^{k}e_{j}^{k}=\delta_{ij},\
S_{i}=e_{i}^{j}S^{j}.\nn 
\end{eqnarray}
The space of states is a direct product of three components:

\vsp\paragraph*{Space of functions $\Psi(P)$} 
with the definition $Z=-i\df/\df P$.

\vsp\paragraph*{Fock space} with a vacuum 

\begin{eqnarray}
&&~\hspace{-10mm}
a_{k}\ket{0}=0,\ k>0,\quad a_{k}^{+}\ket{0}=0,\ k<0\label{vacuum}
\end{eqnarray}
and states created from vacuum by operators

\begin{eqnarray}
&&\ket{\{n_{k}\}}=
\prod_{k>0,k\neq s}{{{1}\over{\sqrt{k^{n_{k}}n_{k}!}}}}
(a_{k}^{+})^{n_{k}}\cdot\nn\\
&&\cdot\prod_{k<0,k\neq-s}{{{1}\over{\sqrt{(-k)^{n_{k}}n_{k}!}}}}
(a_{k})^{n_{k}}
\ket{0}.\nn
\end{eqnarray}

For instance, we will write 
$\ket{1_{1}2_{-1}}={\textstyle{1\over\sqrt{2}}}a_{1}^{+}a_{-1}^{2}\ket{0}$
etc. So defined state vectors have positive norm. Occupation numbers
\begin{eqnarray}
&&~\hspace{-8mm}
n_{k}={{1}\over{|k|}}\;:a_{k}^{+}a_{k}:\;={{1}\over{|k|}}\;\cdot\;
  \left\{ \begin{array}{l}
  a_{k}^{+}a_{k},\;k>0\\
   a_{k}a_{k}^{+},\;k<0
  \end{array} \right.=0,1,2...\nn
\end{eqnarray}

\noindent{\it Remark:} one-component oscillator variables $a_{k}$ 
used throughout this paper are related with commonly applied two-component
oscillators $a_{k1,2}$ by the formulae of Section~\ref{gribsec},
which on quantum level become:
$a_{k}=a_{k1}-ia_{k2}$, $a_{k}^{+}=a_{-k1}+ia_{-k2}$.
The inverse formulae are $a_{k1}=(a_{k}+a_{-k}^{+})/2$,
$a_{k2}=i(a_{k}-a_{-k}^{+})/2$.
Here taking {\it two sets} of oscillator variables 
$a_{ki},i=1,2$ with $a_{ki}^{+}=a_{-ki}$ 
we construct {\it one set} without this property.
Usage of one-component oscillators simplifies the algebra.
It's easy to verify that the definition of vacuum (\ref{vacuum})
is equivalent to a standard one $a_{ki}\ket{0}=0,k>0$
and the states $\ket{\{n_{k}\}}$ are the linear combinations
of $\prod_{k>0}(a_{k1}^{+})^{n_{k1}}(a_{k2}^{+})^{n_{k2}}\ket{0}$.

\vsp\paragraph*{Quantum top:} the space of states is formed by functions
$\Psi(e),\ e_{i}^{j}\in SO(3)$. For the rotation group 
two representations are possible: single- and double-valued \cite{Weyl}. 
Spin is defined as differential operator 
$S^{i}=-i\epsilon_{ijk}e_{l}^{j}\partial/\partial e_{l}^{k}$,
while the projection of spin onto the coordinate system $\vec e_{i}$ is
$S_{i}=i\epsilon_{ijk}e^{l}_{j}\partial/\partial e^{l}_{k}$.
Operator $S^{i}$ generates the rotation of the coordinate system 
$\vec e_{i}$ in external space, while $S_{i}$ generates
the rotations about the axes $\vec e_{i}$. These transformations
act on different indices in $e_{j}^{k}$. They commute, therefore
$S^{3}$ and $S_{3}$ are simultaneously observable.
Matrix elements of spin components 
do not depend on the representation of the algebra
and have well known form \cite{LLifsh}: 
\begin{eqnarray}
&&~\hspace{-9mm}
\bra{S(S_{3}-1)S^{3}}S_{+}\ket{SS_{3}S^{3}}=\sqrt{S(S+1)-S_{3}(S_{3}-1)},
\nn\\
&&~\hspace{-9mm}
\bra{S(S_{3}+1)S^{3}}S_{-}\ket{SS_{3}S^{3}}=\sqrt{S(S+1)-S_{3}(S_{3}+1)},
\nn
\end{eqnarray}
all other elements vanish. Concrete representation of quantum top
is described by Wigner's functions \cite{Wigner,LLifsh}:
$$\ket{SS_{3}S^{3}}={\cal D}^{S}_{S_{3}S^{3}}(e),\quad
 S_{3},S^{3}=-S,-S+1\; ...\; S.$$
Here $S$ characterizes the eigenvalue of Casimir
operator $\vec S^{2}=S^{i}S^{i}=S_{i}S_{i}=S(S+1)$, commuting with
all spin components. $S$ is 
integer for single-valued representation of $SO(3)$ 
and half-integer for double-valued one. Further constraint 
$S_{3}=A_{3}\in\Z$ selects only integer $S$-values. 
The paper \cite{partI-III} Part~III explains in more details, 
why in this problem only integer values of $S$ are available.
The constraints, being functions of $S_{i}$ and oscillator variables,
commute with $\vec S^{2}$, as a result, the determination of mass spectrum 
can be performed separately for each $S$ value.

\vsp\paragraph*{Lorentz group generators} are directly defined 
by their classical expressions (\ref{Xdef}): 
\begin{eqnarray}
&&M_{\mu\nu}=X_{[\mu}P_{\nu]}+\epsilon_{ijk}N_{\mu}^{i}N_{\nu}^{j}S^{k},\nn
\end{eqnarray}
where the square brackets denote antisymmetrization of indices.
The generators are Hermitian operators, forming closed algebra 
of Lorentz group, see Appendix~2. Acting on the state vectors,
the second term in this expression generates the rotations of $\vec e_{i}$
in the argument of the wave function, while the first term generates
Lorentz transformations of $P_{\mu}$, associated changes of CMF
axes $N^{\alpha}_{\mu}$ and certain rotations of the basis 
$\vec e_{i}$ in CMF.

\vsp\paragraph*{Constraints} ~\hspace{-5mm}
$\chi_{3}\ket{\Psi}=H\ket{\Psi}=0$
are imposed~on the states according to Dirac's theory
of constra\-ined dynamical systems \cite{Dirac}. 
Here $\chi_{3}=S_{3}-A_{3}^{(s)}$,

\begin{eqnarray}
&&A_{3}^{(s)}=
\sum_{k\neq 0,\pm s}{\textstyle{1\over k}}:a_{k}^{+}a_{k}:\;=
\sum_{k\neq 0,\pm s}\sgn k\cdot n_{k},\nn\\
&&L_{0}^{(s)}=
\sum_{k\neq 0,\pm s}:a_{k}^{+}a_{k}:\;=
\sum_{k\neq 0,\pm s}|k|\cdot n_{k},\nn
\end{eqnarray}
and the Hamiltonian $H$ is a mass shell condition constructed below.
The operators $\chi_{3}$ and $H$ are Hermitian.
For the operators entering in Hamiltonian: 
$d_{\pm},f_{\pm},g_{\pm},\Sigma_{\pm}$ 
and their conjugates we reserve a term {\it elementary operators}. 
They can be defined by their classical expressions (\ref{defelem}),
with only replacement of complex conjugation to Hermitian one.
These definitions have no ordering ambiguities. 
For the states with the finite number of occupied modes 
the matrix elements $\bra{\{n'_{k}\}}op\ket{\{n_{k}\}}$
of elementary operators $op$ are described by finite sums.

\vsp\paragraph*{Symmetries} $R_{3},D_{2s}$ 
in quantum theory are generated by operators 
$$R_{3}=e^{-i\chi_{3}\alpha},\ 
\tilde D_{2s}=e^{iL_{0}^{(s)}\pi/s},$$
where tilde indicates that the symmetry acts in the space
of internal variables, not including the mean coordinate $X$
(due to omission of $P^{2}$ in the generator).
It's convenient to introduce the notion of 
$(\Delta S_{3},\Delta A_{3},\Delta L_{0})$-{\it charges}
for operators satisfying the relations:
$[S_{3},op]=\Delta S_{3}\cdot op$,
$[A_{3}^{(s)},op]=\Delta A_{3}\cdot op$,
$[L_{0}^{(s)},op]=\Delta L_{0}\cdot op$,
i.e. the operators, which increase or decrease the quantum numbers
$(S_{3},A_{3}^{(s)},L_{0}^{(s)})$ by 
$(\Delta S_{3},\Delta A_{3},\Delta L_{0})$ units:
$$op\ket{S_{3},A_{3}^{(s)},L_{0}^{(s)}}=
\ket{S_{3}+\Delta S_{3},A_{3}^{(s)}+\Delta A_{3},L_{0}^{(s)}+\Delta L_{0}}.$$
Particularly, $\Delta S_{3}(a_{n})=0$, 
$\Delta A_{3}(a_{n})=-1$, $\Delta L_{0}(a_{n})=-n$.
Hermitian conjugation of operators reverses the sign of 
their charges. The elementary operators have $\Delta S_{3}=0$ and 
definite $(\Delta A_{3},\Delta L_{0})$-charges, see Table~\ref{tabord}.
The discrete symmetry $D_{2s}$ is also characterized by the charge 
$Q=L_{0}^{(s)}\mod 2s$, so that the subspaces with given $Q$
are eigenspaces $\tilde D_{2s}\ket{Q}=e^{i\pi Q/s}\ket{Q}$, and
$D_{2s}$-symmetric operators keep these subspaces invariant.

For spin components the commutation relations (\ref{canon}) 
correspond to raising/lowering operators
\begin{eqnarray}
&&~\hspace{-8mm}
S^{\pm}=S^{1}\pm iS^{2},\ 
[S^{3},S^{\pm}]=\pm S^{\pm},\ [S^{+},S^{-}]=2S^{3},\nn\\
&&~\hspace{-8mm}
S_{\pm}=S_{1}\pm iS_{2},\ 
[S_{3},S_{\pm}]=\mp S_{\pm},\ [S_{+},S_{-}]=-2S_{3},\nn
\end{eqnarray}
i.e. $\Delta S^{3}(S^{\pm})=\pm1$, $\Delta S_{3}(S_{\pm})=\mp1$.
In further mechanics only low-index operators $S_{i}$ will participate. 
Linear combinations $l_{\pm}=\lambda_{1}S_{\pm}+\lambda_{2}\Sigma_{\pm}$
do not have definite $\Delta S_{3}$ and $\Delta A_{3}$ charges,
but have definite $\Delta\chi_{3}=\Delta S_{3}-\Delta A_{3}=\mp1$.
Because the classical symmetries are defined by polynomial structure, 
they are preserved on the quantum level. 
Particularly, quantum Hamiltonian possesses $R_{3},D_{2s}$-symmetries,
as a result, its non-zero matrix elements form blocks, located on 
$(\chi_{3}Q)$-diagonal: $\bra{\chi_{3}=0,Q}H\ket{\chi_{3}=0,Q}$.

\vsp\paragraph*{Quantum Hamiltonian}~

\vspace{2mm}
\noindent Further we fix $s=2$. Quantum analog of (\ref{alp2cl}) 
is constructed as follows:
\begin{eqnarray}
&&\alpha_{2}=\sum\limits_{n\geq1}\tilde n_{1}^{-2n+1}:P_{n}:,\label{alp2q}
\end{eqnarray}
where $\tilde n_{1}=a_{1}^{+}a_{1}+c_{1}$. Here we introduce a constant term 
$c_{1}$, whose contribution vanishes on classical level
(in the limit of large occupation numbers $a_{1}^{+}a_{1}$),  
and add analogous terms in quantum definition 
$\gamma=(L_{0}^{(2)}+c_{2})^{-1/2}$ and in the definition
of mass shell condition, which we fix as follows:
\begin{eqnarray}
&&{P^{2}\over2\pi}=L_{0}^{(2)}+2\alpha_{2}\alpha_{2}^{+}+c_{3}.\label{P2q}
\end{eqnarray}
The operator $\alpha_{2}$ has charges $(\Delta\chi_{3}=-1, \Delta Q=0)$,
while for $P^{2}$ $(\Delta\chi_{3}=0, \Delta Q=0)$.
Polynomials $P_{n}$ are defined by expressions of Appendix~3,
where we substitute the definitions (\ref{defelem}) of elementary operators
and fix the ordering, shown in Table~\ref{tabord}.
This ordering puts $L_{0}^{(2)}$-lowering elementary operators
to the right from $L_{0}^{(2)}$-raising ones, thus providing better
convergence properties for the expansion (\ref{alp2q}). Particularly, 
the matrix elements of $\alpha_{2}$ between the states with finite 
$L_{0}^{(2)}$ are given by finite sums, and large $n$ terms
of (\ref{alp2q}) contribute only to the matrix elements with large 
$L_{0}^{(2)}$, due to the following property \cite{partI-III}:

\vspace{1mm}\noindent
$\bra{L_{0}^{(2)}=N_{1},S} :P_{n}: \ket{L_{0}^{(2)}=N_{2},S}=0$, if \\
$n>\min\{1+(N_{1}+N_{2})/2,\ (4+2(N_{1}+N_{2})+4S)/5\}$.

\def\figpage{
\begin{table}\label{tabord}
\begin{center}
\tabnum: $L_{0}^{(2)}$-normally ordered elementary operators

~

\vspace{-10mm}

$$\begin{array}{|c|cccccccccccccccc|}\hline
&a_{1}^{+}&(d_{-})^{+}&d_{+}&(f_{-})^{+}&f_{+}&(g_{-}')^{+}&g_{+}&
\Sigma_{-}&\Sigma_{+}&(g_{+})^{+}&g_{-}'&
(f_{+})^{+}&f_{-}&(d_{+})^{+}&d_{-}&a_{1}\\\hline
\Delta L_{0}^{(2)}&1&4&4&2&2&2&2&0&0&-2&-2&-2&-2&-4&-4&-1\\
\Delta A_{3}^{(2)}&1&-1&1&0&0&2&-2&-1&1&2&-2&0&0&-1&1&-1\\\hline
\end{array}
$$
\end{center}
\end{table}
}\twocolumn[\figpage]

~

\vspace{-7mm}\noindent

\vspace{1mm}
In the products of spin components
$S_{\pm}$, commuting with elementary operators, we select the
ordering $S_{+}S_{-}$, so that $S_{3}$-raising operator
$S_{-}$ stands on the right and annulates the states
with maximal spin projection $S_{3}=S$. We remind that
classically $S_{3}=S=P^{2}/2\pi$ corresponds to one-modal
solution, straight-line string, associated with the leading 
Regge trajectory. 

The introduced ordering possesses the other feature 
convenient for computations: the matrix elements of 
the normally ordered operators restricted to
finite-dimensional subspaces, behave regularly when
the dimension of subspaces increases. Namely, 
the matrix in larger space includes the matrix in
smaller space as an exact submatrix. 
Let \hbox{$op_{i}(N)=\bra{L_{0}^{(2)}\leq N,S}op_{i}\ket{L_{0}^{(2)}\leq N,S}$}
be restrictions of elementary operators $op_{i}$ to finite-dimensional
space $L_{0}^{(2)}\leq N$, at given $S$. Let \hbox{$:P(op_{i}):$}
be $L_{0}^{(2)}$-normally ordered polynomial of $op_{i}$. Then
(i) at \hbox{$N_{1}<N_{2}$} the matrix \hbox{$:P(op_{i}(N_{1})):$} is
a submatrix of \hbox{$:P(op_{i}(N_{2})):$} and 
(ii) 
\hbox{$:P(op_{i}(N)):=$}
\hbox{$\bra{L_{0}^{(2)}\leq N,S}:P(op_{i}):
\ket{L_{0}^{(2)}\leq N,S}$.}
Without normal ordering
this simple behavior would be violated.

Practically, we compute the matrix elements of elementary operators 
in $(L_{0}^{(2)}\leq N,S)$-subspace.
The dimension of this subspace is rapidly increasing
with $N,S$, e.g. $dim=281216$ for $N=20$, $S=6$, 
so that the elementary operators are represented 
as matrices of very large size ($dim\times dim$).
The matrices have noticeable block structure,
corresponding to their $(\Delta S_{3},\Delta A_{3},\Delta L_{0})$-charge 
properties. As a result, non-zero matrix elements of $\alpha_{2}$, 
necessary for computation of $P^{2}$, are located in the blocks 
$\bra{S_{3}=A_{3}^{(2)},Q}\alpha_{2}\ket{S_{3}=A_{3}^{(2)}+1,Q}$, 
while $P^{2}$ itself is located in 
$\bra{S_{3}=A_{3}^{(2)},Q}P^{2}\ket{S_{3}=A_{3}^{(2)},Q}$.

\begin{figure}\label{elka-corrected}
\begin{center}
~\epsfxsize=5cm\epsfysize=7.5cm\epsffile{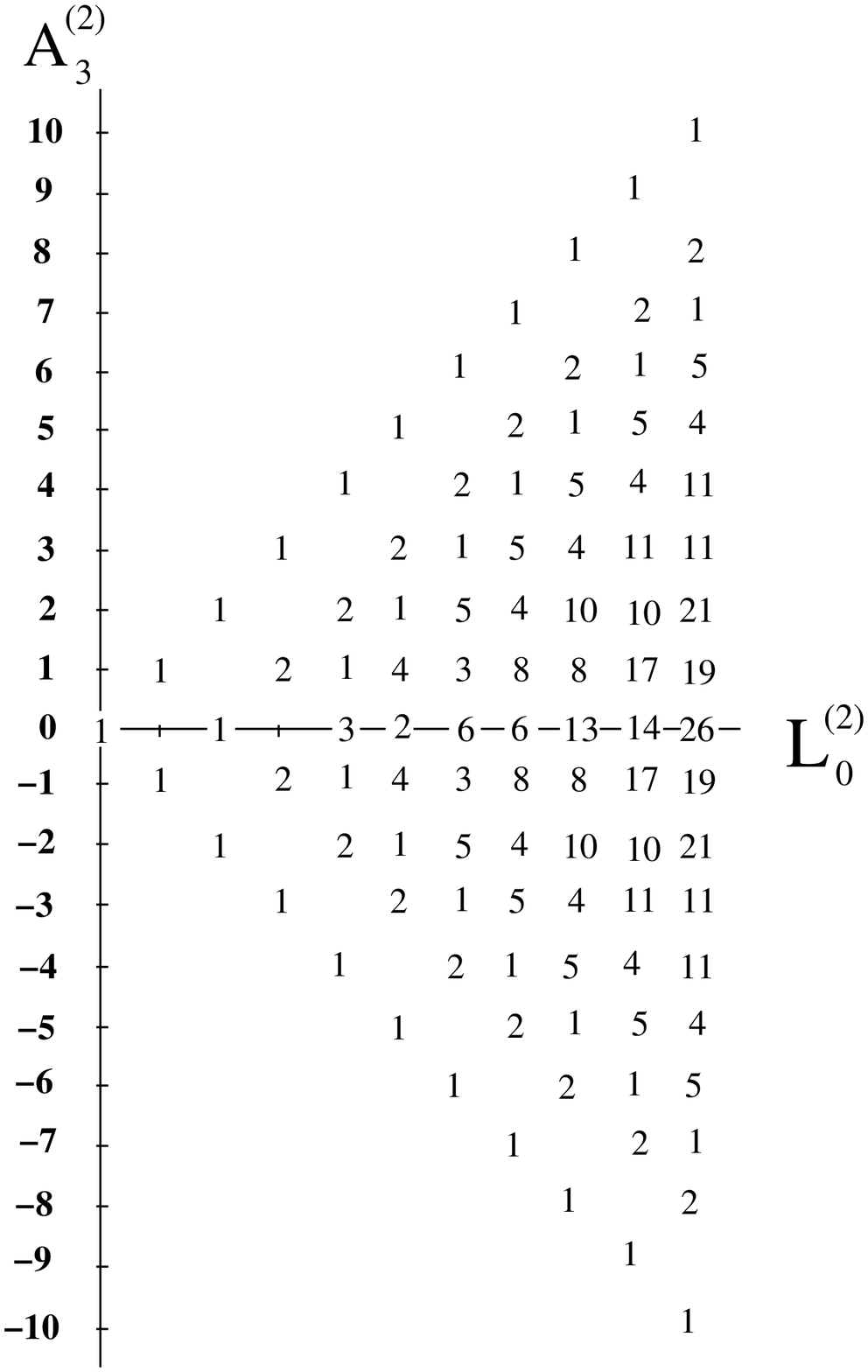}

\fignum Spectrum $(L_{0}^{(2)},A_{3}^{(2)})$.
\end{center}
\end{figure}

\begin{figure}\label{spec}
\begin{center}
~\epsfxsize=8cm\epsfysize=6cm\epsffile{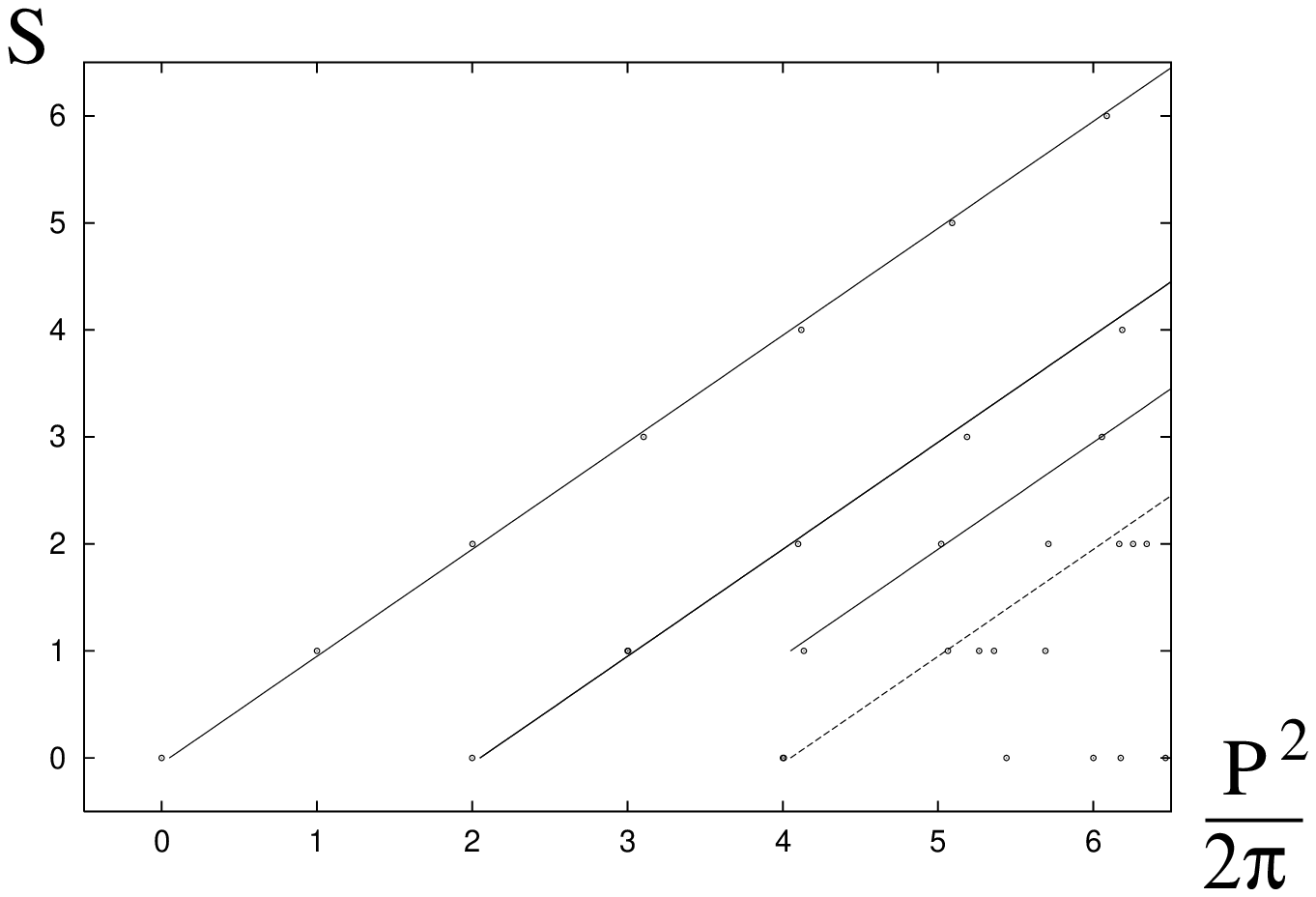}

\fignum Spectrum $(P^{2}/2\pi,S)$.
\end{center}
\end{figure}

In addition to these properties we use the fact that elementary matrices 
inside $(S_{3},A_{3}^{(2)},L_{0}^{(2)})$-blocks are very sparse
(at large $N,S$ their non-zero content occupies less than 1\% of the blocks),
and implement special algorithms for sparse block matrix computations,
described in more details in \cite{partI-III}. Finally, we determine the
spectrum of $P^{2}/2\pi$ up to the values $N=20$, $S=6$ and 
the number of terms in expansion (\ref{alp2q}) up to $n=3$.
The resulting spectrum $(P^{2}/2\pi,S)$ is shown on \fref{spec}.
The spectrum has common features with the upper part
of $(L_{0}^{(2)},A_{3}^{(2)})$ spectrum, 
shown on \fref{elka-corrected}. 
The beginning of the spectrum $(P^{2}/2\pi,S)$ 
consists of three almost linear Regge trajectories. There is a 2-unit 
gap between the first and the second trajectories. The third trajectory
starts at $S=1$ level. For the next trajectories the degenerate states of
\fref{elka-corrected} become splitted on \fref{spec}.
The states at $(P^{2}/2\pi,S)=(3,1)$ and $(4,0)$
comprise two numerically close pairs with $P^{2}/2\pi=3,3.0046$
and $P^{2}/2\pi=4,4.0066$, while the other states on \fref{spec}
are non-degenerate (not counting trivial degeneracy for 
the upper-index $S^{3}=-S...S$ and direction of $P_{\mu}$). 
The spectrum is computed for the values
$c_{1}=2,\ c_{2}=4,\ c_{3}=0$. Smaller values of $c_{1},c_{2}$
correspond to higher non-linearities in the spectrum, while
larger values of $c_{1},c_{2}$ make the spectrum more linear
and closer to the spectrum of $(L_{0}^{(2)},A_{3}^{(2)}\geq0)$.
Further for clarity we fix $c_{3}=0$. 

\vspace{5mm}
\begin{table}\label{tabZ}
\tabnum: eigenvectors with $P^{2}/2\pi\in\Z$.

\vspace{-5mm}
$$
\begin{array}{|c|c|c|}\hline
P^{2}/2\pi=L_{0}^{(2)}&S=S_{3}=A_{3}^{(2)}&\ket{\{n_{k}\}}\\\hline
0&0&\ket{0}\\ 
2&0&\ket{1_{1}1_{-1}}\\ 
4&0&\ket{2_{1}2_{-1}}\\
1&1&\ket{1_{1}}\\ 
3&1&\ket{1_{3}}\\\hline
\end{array}
$$

\end{table}

\vspace{3mm}\noindent
The following properties
of the spectrum have been found in \cite{partI-III}:

\vspace{0.5mm}\noindent
1) for all $c_{1,2}>0$ the states from Table~\ref{tabZ} are annulated by
$\alpha_{2}^{+}$ and have integer-valued 
$P^{2}/2\pi$.

\vspace{0.5mm}\noindent
2) let's keep in $\alpha_{2}^{+}$ only 
the leading $(1/\tilde n_{1})$-term:
$\alpha_{2}^{+}|_{n=1}=(-\Sigma_{-}+S_{-}/\gamma)/\tilde n_{1}$. 
For all $c_{1,2}>0$ 
the states from the first two Regge-trajectories
are annulated by $\alpha_{2}^{+}|_{n=1}$. In this approximation
the first two Regge-trajectories have integer-valued $P^{2}/2\pi$
and are linear: $P^{2}/2\pi=S+k$, $k=0,2$.

\vspace{0.5mm}\noindent
3) in the limit $1\ll c_{1}^{2}\ll c_{2}\ll c_{1}^{4}$ 
the spectrum of $(P^{2}/2\pi,S)$ tends to the spectrum of
\hbox{$(L_{0}^{(2)},A_{3}^{(2)}\geq0)$}.

\vspace{2mm}
\noindent{\it Remark}:
the first Regge trajectory corresponds to the straight-line string's
northern pole solution (\ref{1mode}), further trajectories are created
by expansion (\ref{a2cl}) in the vicinity of the northern pole solution.
Classically there are gauge equivalent solutions near the southern pole
and the equator, which possess the same $(P^{2},S)$. In quantum
mechanics we do not see these additional solutions in the spectrum.
Particularly, the first Regge trajectory is non-degenerate.
We conclude, that the equivalence between these solutions is lost
in quantum mechanics. The spectra for additional solutions
can be shifted to the region of large masses, or the quantum expansions
(\ref{alp2q}) can even diverge on these solutions. Indeed, the classical 
solutions in the vicinity of the northern pole (\ref{1mode}) possess
large $n_{1}$ and small $n_{k}$, $k\in\Z\backslash\{0,1,\pm2\}$,
providing the convergence for $(1/n_{1})$-expansion (\ref{a2cl}).
In quantum mechanics the convergence of expansion (\ref{alp2q})
is supported by the finite number of occupied modes in 
$(L_{0}^{(2)}\leq N)$-spaces and the normal ordering of operators.
For the solutions near the southern pole $n_{1}\to0$, and the usage
of $(1/n_{1})$-expansions is problematic. One can use $(1/n_{-1})$-series
to construct a definition of mass shell condition, alternative to
(\ref{P2q}), however these definitions will substantially differ 
on the quantum level and in fact will create two distinct theories.
For the solutions on the equator infinite number of oscillator modes
are excited, and for these solutions the convergence of the expansions
is not guaranteed neither on classical nor on quantum level.
This argumentation explains why the usage of $(1/n_{1})$-series
in the vicinity of the northern pole solution preserves only one
Gribov's copy in the quantum theory.

\vspace{-2mm}
\section{Discussion}

\vspace{-2mm}
In this paper we have constructed the quantum theory of 
open Nambu-Goto string in the space-time of dimension $d=4$. 
The general approach is the selection of the light-cone gauge 
with the gauge axis related in Lorentz-invariant way with the world sheet.
In this approach the Lorentz group transforms the world sheet 
together with the gauge axis and is not 
followed by reparametrizations. As a result, the theory
becomes free of anomalies in Lorentz group and in the group
of internal symmetries of the system. The constructed
quantum theory possesses spin-mass spectrum with Regge-like
behavior.

Certain problems are still present in this theory,
which however do not hinder its implementations, 
e.g. for the construction of string models of the hadrons.
The results, produced by the theory, are influenced
by ordering of operators and other details of quantization procedure.
The theory does not contain {\it algebraic anomalies}, but possesses 
the features, which can be called {\it spectral anomalies}.
Particularly, Hamiltonian $P^{2}/2\pi$, classically generating
$2\pi$-periodic evolution, in quantum theory is influenced
by ordering rules and does not have strictly equidistant spectrum. 
This fact does not create problems for the hadronic models,
where this spectrum is subjected to phenomenological corrections
and experimentally is not strictly equidistant as well. 
The theory also 
possesses a topological defect, appearing as a discrete gauge
symmetry, identifying the points in the phase space (Gribov's copies).
This classical symmetry is related with discrete non-linear 
reparametrizations of the world sheet and is not preserved
on the quantum level. In our construction we use the expansion series
in the vicinity of one Gribov's copy, by these means distinguishing 
it in the quantum theory. We have also shown that
the leading term of the expansion, which has a minimal
ordering ambiguity and is easier for computation, is sufficient 
to reproduce Regge behavior of the spectrum. Therefore,
practically one can keep this term to describe the main effect
and include further terms in the form of phenomenological corrections,
together with the contributions of other nature 
\cite{NambuHiggs,chrom1,chrom2,Artru,thickness,
straight-em1,3str,3strSharov,BarbNest}: 
gluonic tube thickness, quark masses and charges,
spin-orbital interaction, etc.

In conclusion we perform the comparison of
the obtained spectrum with the results of other
existing approaches to non-critical quantization
of string theory. This problem was previously solved for 
certain submanifolds in the phase space of open string
at $d=4$, which represent the world sheets of a special form, 
i.e. particular types of string motion.

\begin{figure}\label{fspec0}
\begin{center}
~\epsfxsize=6cm\epsfysize=4.5cm\epsffile{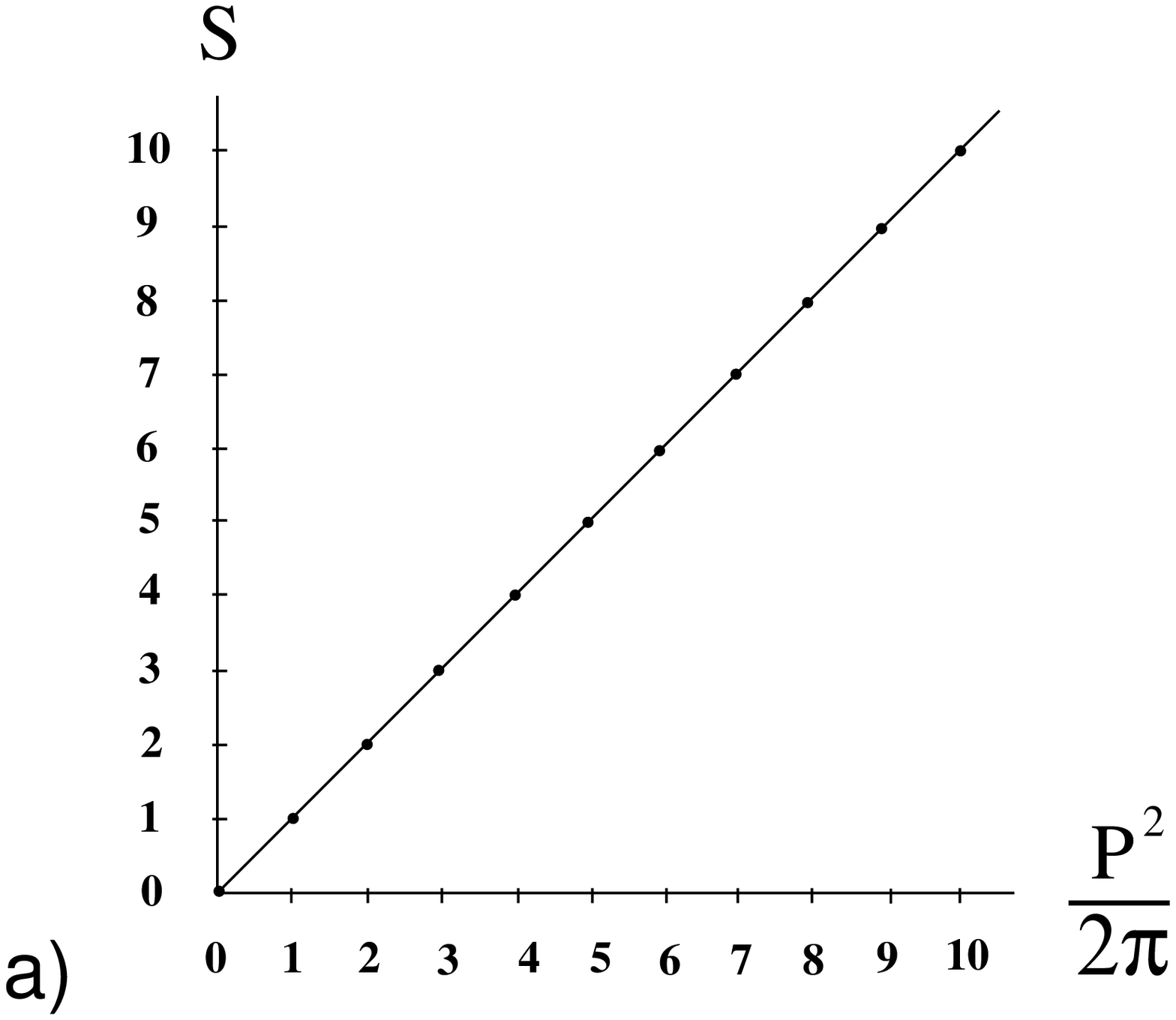}

~\epsfxsize=6cm\epsfysize=4.5cm\epsffile{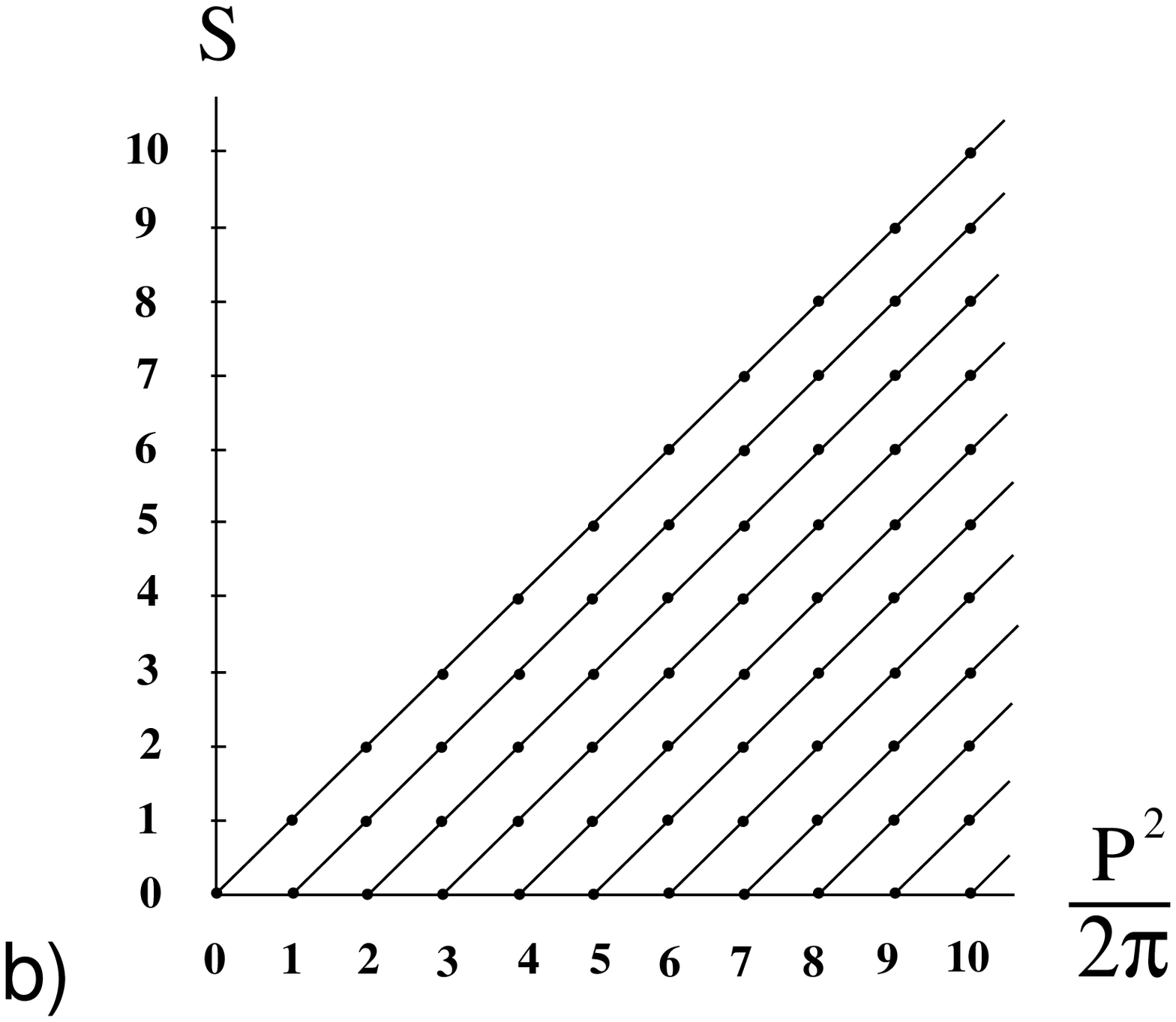}

~\epsfxsize=6cm\epsfysize=4.5cm\epsffile{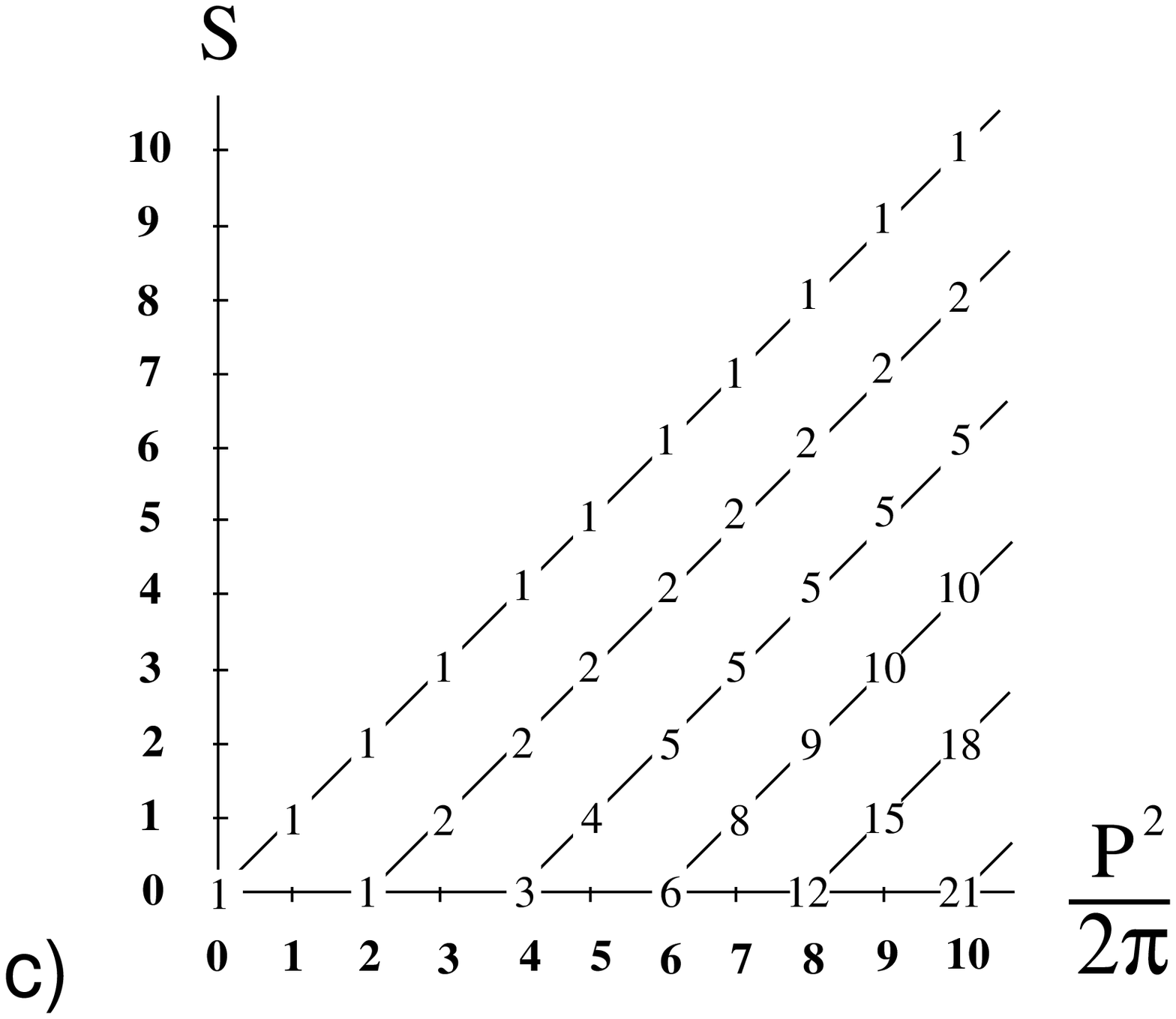}

~\epsfxsize=6cm\epsfysize=4.5cm\epsffile{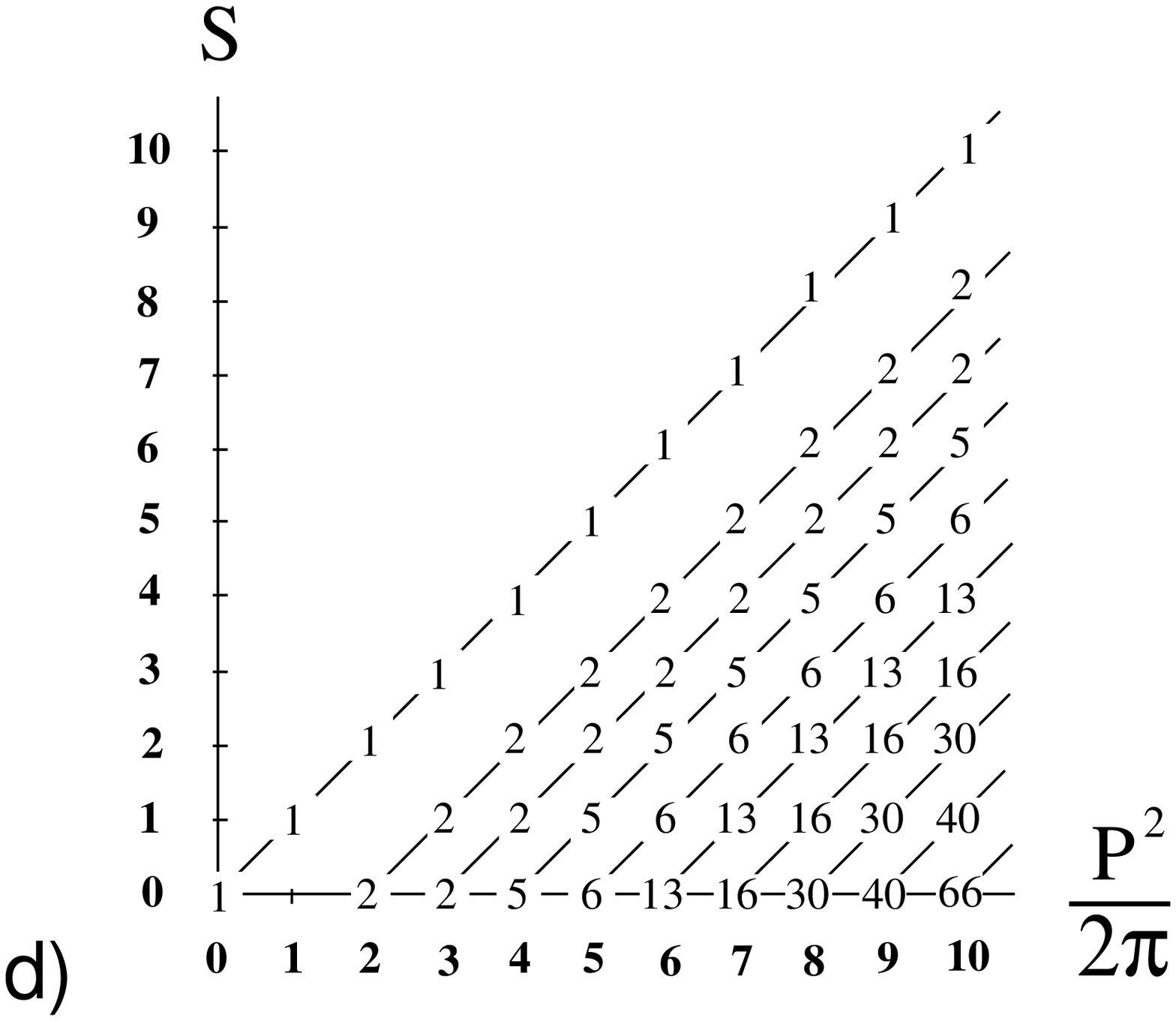}

\end{center}

\fignum~Spin-mass spectra for the theories of 
restricted types of string motion,
admitting anomaly free quantization at $d=4$:
(a)~straight-line string~\cite{slstring},
(b)~2-parametric string~\cite{2par},
(c)~axially symmetric string~\cite{ax},
(d)~stratum $\nu_{0}=1$ in the phase space~\cite{zone}.

\end{figure}

The first example is given by the above mentioned 
straight-line string solution \cite{slstring},
whose spin-mass spectrum consists of a single leading 
Regge trajectory, see \fref{fspec0}a. 
In \cite{straight-em1} this quantum model
was extended by the spin and electric charges 
of the quarks, describing well the experimentally 
observed states in the spectrum of light mesons,
and some of their radiative transitions. 

The straight-line string can be considered as essentially
one-parametric solution, where the only parameter is the 
length of the string. The paper \cite{2par} considered 
a generalization of this mechanics, and shown that
an arbitrary two-parametric family of string motion, 
containing the straight-line string solution as a subset,
after a definite weak topological restriction,
admits anomaly free quantization.
Its spectrum, presented at \fref{fspec0}b, 
contains infinite number of Regge trajectories.
All the states in this spectrum are non-degenerate
(have multiplicity one).

\newpage

\vspace{5mm}

The paper \cite{ax} has considered a special type
of string motion, where the string in the center-of-mass
frame possesses an axial symmetry. This type of motion
already includes an infinite number of degrees of freedom.
It allows anomaly free quantization in Lorentz-invariant 
light cone gauge, related with the axis of symmetry.
Corresponding spectrum, shown at \fref{fspec0}c, 
consists of the infinite number of Regge trajectories
with non-trivial multiplicities. 

The paper \cite{zone} has considered the motions
of a general form, and obtained spin-mass spectrum
for a special infinite-dimensional subset ($\nu_{0}=1$ stratum), 
possessing a simple topological structure.
Its spectrum (\fref{fspec0}d) as well contains
the infinite number of Regge trajectories, with 
the multiplicities constant along the trajectories,
and 2-unit gap between the first two trajectories.

The presented spectra possess evident structural similarities.
The discrepancies are caused by the fact that the
approaches \cite{slstring,2par,ax,zone}
consider different types of motion, and as well
by deviations $\sim\hbar$ inherent to quantization procedure.
Conceptually, these theories are different implementations of 
the same underlying ideas, leading to the main common result:
the absence of quantum anomalies in Lorentz group 
at physical dimension of the space-time.

\vspace{-2mm}
\paragraph*{Appendix~1:}~\\
\hbox{symplectic structure of the phase space.}

\vspace{2mm}
\noindent 
The formalism of symplectic geometry \cite{slstring,Arnold-mat-phys}
is convenient to describe the structure of the phase space
in Hamiltonian dynamics with constraints. The phase space
is a smooth manifold, endowed by a closed non-degenerate 
differential 2-form $\Omega = {{1}\over{2}}\omega_{ij}dX^{i}\wedge dX^{j}$ 
(in some local coordinates $X^{i},\  i=1,\ldots,2n$). 
Poisson brackets are defined by the form as
$\{X^{i},X^{j}\}=\omega^{ij}$, where $\|\omega^{ij}\|$ is inverse to 
$\|\omega_{ij}\|$: $\omega_{ij}\omega^{jk}=\delta_{i}^{k}$.
Let's consider a surface in the phase space, given by the 2nd class 
constraints: 
$\chi_{\alpha}(X)=0\ (\alpha =1,\ldots,r),$ $det\|\{\chi_{\alpha},
\chi_{\beta}\}\|\neq 0 $. Reduction on this surface consists in the 
substitution of its explicit parametrization 
$X^{i}=X^{i}(u^{a})\ (a=1,\ldots,2n-r)$ into the form: 
\begin{eqnarray}
&&\Omega = {{1}\over{2}}\Omega_{ab}du^{a}\wedge du^{b},\quad
 \Omega_{ab}={{\partial X^{i}}\over{\partial u^{a}}}\omega_{ij}
 {{\partial X^{j}}\over{\partial u^{b}}}.\nn
\end{eqnarray}
Here $det\|\Omega_{ab}\|\neq 0$.
Matrix $\|\Omega^{ab}\|$ , inverse to 
$\|\Omega_{ab}\|$, defines Poisson brackets on the surface:
 $\{u^{a},u^{b}\}=\Omega^{ab}$.
This method is equivalent to commonly used Dirac brackets' 
formalism \cite{Dirac}:
$$ \{u^{a},u^{b}\}^{D}=\{u^{a},u^{b}\} - \{u^{a},\chi_{\alpha}\}
\Pi^{\alpha\beta}\{\chi_{\beta},u^{b}\}=\Omega^{ab},$$
where $\|\Pi^{\alpha\beta}\|$ is inverse to $\|\Pi_{\alpha\beta}\|$: 
$\Pi_{\alpha\beta}=\{\chi_{\alpha},\chi_{\beta}\}$.
In string theory canonical Poisson brackets 
$\{x_{\mu}(\s),p_{\nu}(\tilde\s)\}=g_{\mu\nu}\delta(\s-\tilde\s)$
correspond to symplectic form $\Omega=\int_{0}^{\pi}d\s\;
\delta p_{\mu}(\s)\wedge \delta x_{\mu}(\s)$.
By a substitution of expressions for $x_{\mu}(\s),p_{\mu}(\s)$
in terms of $Q_{\mu}(\s)$, given by eqs.(\ref{Qxp}),
it is transformed to the form
\begin{eqnarray}
&&\Omega=\half\; dP_{\mu}\wedge dQ_{\mu}(0)+\nn\\
&&\quad
+\quart\intt\;\delta Q'_{\mu}(\s)\wedge\delta Q_{\mu}(\s),\nn
\end{eqnarray}
and by substitution of light cone parametrization~(\ref{Qdlcg}):
\begin{eqnarray}
&&~\hspace{-1.2cm}
\Omega=dP_{\mu}\wedge dZ_{\mu}+
 \sum_{k \neq 0}\frac{1}{ik}\; da_{k}^{*}\wedge da_{k}+\nn\\
&&+\half d{\vec e_{i}} \wedge d({\vec S}\times{\vec e_{i}}).\label{symform}
\end{eqnarray}
Inverting the coefficient matrix of this form (in the presence
of orthonormality constraints $\vec e_{i}\vec e_{j}=\delta_{ij}$),
we obtain Poisson brackets (\ref{Pb1}). Further,
substituting the gauge fixing conditions (\ref{gauge}) to the form 
(\ref{symform}), we see that $a_{\pm s}$-terms cancel each other:
${\textstyle{1\over is}}(da_{s}^{*}\wedge da_{s}-da_{-s}^{*}\wedge da_{-s})=0$.
After the reduction we obtain the same canonical basis as (\ref{Pb1}),
but with $\{a_{k},a_{n}^{*}\}=ik\delta_{kn},\ k,n\in\Z\backslash\{0,\pm s\}$.
To study this property in more detail, we introduce the variables 
$q_{1}=\mbox{Re}(a_{s}+a_{-s})/2,\ q_{2}=-\mbox{Im}(a_{s}+a_{-s})/2,\ 
p_{1}=\mbox{Im}(a_{s}-a_{-s})/2,\ p_{2}=\mbox{Re}(a_{s}-a_{-s})/2$, 
and rewrite $a_{\pm s}$-terms of the symplectic form
as ${\textstyle{4\over s}}(dp_{1}\wedge dq_{1}-dp_{2}\wedge dq_{2})$.
Gauge fixing conditions (\ref{gauge}) are rewritten to $q_{1}=q_{2}=0$,
now we see that the variables $p_{1,2}$, canonically conjugated 
to $q_{1,2}$, drop out from the symplectic form. 
$p_{1,2}$ should be expressed in terms of other dynamical variables
from $\chi$-constraints and, independently on the complexity
of these expressions, Poisson brackets of the mechanics remain simple. 

\vspace{-2mm}
\paragraph*{Appendix~2:}~\\ 
absence of anomalies in $[M_{\mu\nu},M_{\rho\s}]$.

\vspace{2mm}\noindent
Representing $M_{\mu\nu}=Z_{[\mu}P_{\nu]}-\half\epsilon_{ijk}
\Gamma^{ij}_{[\mu}P_{\nu]}S^{k}+\epsilon_{ijk}N_{\mu}^{i}N_{\nu}^{j}S^{k}=
Z_{[\mu}P_{\nu]}+G^{k}_{\mu\nu}(P)S^{k}$ and writing the
commutator $[Z_{[\mu}P_{\nu]}+G^{k}_{\mu\nu}(P)S^{k},
Z_{[\rho}P_{\s]}+G^{n}_{\rho\s}(P)S^{n}]$,
we see that the term $[Z_{[\mu}P_{\nu]},Z_{[\rho}P_{\s]}]$
coincides with the commutator of Lorentz generators
in a theory of point-like particles, which is free
of any anomalies; in the term $[G^{k}_{\mu\nu}(P)S^{k},
G^{n}_{\rho\s}(P)S^{n}]$ the only non-commuting variables are
$[S^{k},S^{n}]=i\epsilon_{knm}S^{m}$, as a result we obtain
the expression of the form $GGS$, including only commuting 
variables, and the result coincides (up to the multiplier $i$)
with the Poisson bracket of the corresponding classical variables;
in the term $[Z_{[\mu}P_{\nu]},G^{n}_{\rho\s}(P)S^{n}]$
and the remaining one, related with this term by the replacements 
$(\mu\leftrightarrow\rho,\nu\leftrightarrow\s,n\leftrightarrow k)$,
the only non-commuting variables are $[Z_{\mu},G^{n}_{\rho\s}(P)]$,
and the computation of this commutator gives the derivative
$\df G^{n}_{\rho\s}(P)/\df P_{\mu}$, thus we obtain the expression
$(\df G)PS$, which includes the commuting 
operators and again coincides with the Poisson bracket.
Therefore, considered commutator has no anomalous terms
and its computation actually coincides with the computation
of the corresponding Poisson bracket. Note that this bracket
necessarily represents canonical relations for the Lorentz algebra:
\begin{eqnarray}
&&\{M_{\mu\nu},M_{\rho\s}\}=g_{\rho[\nu}M_{\s\mu]}
+g_{\s[\nu}M_{\mu]\rho},\nn
\end{eqnarray}
because the transformations we performed consist
only in the change of canonical basis $a_{i}\to b_{i}$ 
in the phase space, which preserves the original Poisson bracket 
$\{M_{\mu\nu},M_{\rho\s}\}$. This fact was also proven by direct 
computation in \cite{slstring}.

\vspace{-2mm}
\paragraph*{Appendix~3:} coefficients of expansion (\ref{a2cl}).


\baselineskip=0.4\normalbaselineskip\footnotesize

\def\myfrac#1#2{#1/#2}

\vspace{3mm}\noindent
$P_{1}=-{\Sigma_{+}} + \myfrac{{S_{+}}}{{\gamma}},$

\vspace{2mm}\noindent
$P_{2}={{a_{1}^{*}}}^2\,{f^{*}}\,{\Sigma_{-}} + 
  {{a_{1}}}^2\,{{g'}^{*}}\,{\Sigma_{+}} - 
  \myfrac{{{a_{1}^{*}}}^2\,{f^{*}}\,{S_{-}}}
   {{\gamma}} - \myfrac{{{a_{1}}}^2\,{{g'}^{*}}\,
     {S_{+}}}{{\gamma}},$

\vspace{2mm}\noindent
$P_{3}=- {{a_{1}^{*}}}^4\,{f^{*}}\,{g'}\,
     {\Sigma_{-}}   - 
  {{a_{1}}}^2\,{{a_{1}^{*}}}^2\,{f^{*}}\,{{g'}^{*}}\,
   {\Sigma_{-}} - {{a_{1}^{*}}}^4\,{d^{*}}\,
   {{\Sigma_{-}}}^2 $

\noindent
$- {{a_{1}}}^2\,{{a_{1}^{*}}}^2\,f\,
   {f^{*}}\,{\Sigma_{+}} - 
  {{a_{1}}}^4\,{{{g'}^{*}}}^2\,{\Sigma_{+}} - 
  \myfrac{{{a_{1}}}^2\,{{a_{1}^{*}}}^2\,d\,{\Sigma_{-}}\,
     {\Sigma_{+}}}{2} $

\noindent
$+ \myfrac{{{a_{1}^{*}}}^4\,{f^{*}}\,
     {g'}\,{S_{-}}}{{\gamma}} + 
  \myfrac{{{a_{1}}}^2\,{{a_{1}^{*}}}^2\,{f^{*}}\,
     {{g'}^{*}}\,{S_{-}}}{{\gamma}} + 
  \myfrac{2\,{{a_{1}^{*}}}^4\,{d^{*}}\,{\Sigma_{-}}\,
     {S_{-}}}{{\gamma}} $

\noindent
$+ \myfrac{{{a_{1}}}^2\,{{a_{1}^{*}}}^2\,d\,{\Sigma_{+}}\,
     {S_{-}}}{2\,{\gamma}} - 
  \myfrac{{{a_{1}^{*}}}^4\,{d^{*}}\,{{S_{-}}}^2}
   {{{\gamma}}^2} + \myfrac{{{a_{1}}}^2\,{{a_{1}^{*}}}^2\,
     f\,{f^{*}}\,{S_{+}}}{{\gamma}} $

\noindent
$+ \myfrac{{{a_{1}}}^4\,{{{g'}^{*}}}^2\,{S_{+}}}
   {{\gamma}} + \myfrac{{{a_{1}}}^2\,{{a_{1}^{*}}}^2\,d\,
     {\Sigma_{-}}\,{S_{+}}}{2\,{\gamma}} + 
  {{a_{1}}}^2\,{{a_{1}^{*}}}^2\,{\gamma}\,
   {\Sigma_{-}}\,{\Sigma_{+}}\,{S_{+}} $

\noindent
$- \myfrac{{{a_{1}}}^2\,{{a_{1}^{*}}}^2\,d\,{S_{-}}\,
     {S_{+}}}{2\,{{\gamma}}^2} - 
  {{a_{1}}}^2\,{{a_{1}^{*}}}^2\,{\Sigma_{+}}\,
   {S_{-}}\,{S_{+}} - 
  {{a_{1}}}^2\,{{a_{1}^{*}}}^2\,{\Sigma_{-}}\,
   {{S_{+}}}^2 $

\noindent
$+ \myfrac{{{a_{1}}}^2\,{{a_{1}^{*}}}^2\,
     {S_{-}}\,{{S_{+}}}^2}{{\gamma}}.$

\baselineskip=\normalbaselineskip\normalsize


\footnotesize

\def\vsp{\vspace{0mm}}

\end{document}